\documentclass[aps,prd,twocolumn,superscriptaddress,nofootinbib]{revtex4-2}

\usepackage{amsmath,amssymb,graphicx,hyperref,xcolor}
\usepackage{multirow}
\usepackage{physics}
\usepackage{siunitx}
\usepackage{comment}
\usepackage{acronym}
\usepackage[normalem]{ulem}
\usepackage[f]{esvect}
\usepackage{makecell}
\usepackage{orcidlink}
\allowdisplaybreaks[1]

\newcommand{\TRC}{MOE Key Laboratory of TianQin Mission, TianQin Research Center for Gravitational Physics $\&$  School of Physics and Astronomy, Frontiers Science Center for TianQin, CNSA Research Center for Gravitational Waves, Sun Yat-sen University (Zhuhai Campus), Zhuhai 519082, China}
\newcommand{\HUST}{National Gravitation Laboratory, MOE Key Laboratory of Fundamental Physical Quantities Measurements, Department of Astronomy and School of Physics, Huazhong University of Science and Technology, Wuhan 430074, China}
\newcommand{\KIAA}{The Kavli Institute for Astronomy and Astrophysics, Peking University, Beijing 100871, China}
\newcommand{\BIMSA}{Beijing Institute of Mathematical Sciences and Applications, Beijing 101408, China}
\newcommand{\HBU}{ Department of Physics, Hebei Key Laboratory of High-precision Computation and Application of Quantum Field Theory $\&$ Hebei Research Center of the Basic Discipline for Computational Physics, Hebei University,  Baoding, 071002, China}

\hypersetup{
    colorlinks = true,
    linkcolor = blue,
    citecolor = blue,
    urlcolor  = blue
}

\begin{document}

\title{Constructing a gravitational wave analysis pipeline for extremely large mass ratio inspirals}

\author{Tian-Xiao Wang\orcidlink{0009-0006-3173-6506}}
\affiliation{\TRC}
\author{Yan Wang\orcidlink{0000-0001-8990-5700}}
\email{Corresponding author: ywang12@hust.edu}
\affiliation{\HUST}
\author{Alejandro Torres-Orjuela\orcidlink{0000-0002-5467-3505}}
\affiliation{\BIMSA}
\author{Yi-Ren Lin\orcidlink{0000-0003-2509-6558}}
\affiliation{\KIAA}
\author{Hui-Min Fan\orcidlink{0000-0001-9636-6637}}
\affiliation{\HBU}
\author{Ver\'onica V\'azquez-Aceves\orcidlink{0000-0002-9458-8815}}
\affiliation{\KIAA}
\author{Yi-Ming Hu\orcidlink{0000-0002-7869-0174}}
\email{Corresponding author: huyiming@sysu.edu.cn}
\affiliation{\TRC}

\date{\today}

\begin{abstract}
Extremely large mass-ratio inspirals (XMRIs), consisting of a brown dwarf orbiting a supermassive black hole, emit long-lived and nearly monochromatic gravitational waves in the millihertz band and constitute a promising probe of strong-field gravity and black-hole properties.
However, dedicated data-analysis pipelines for XMRI signals have not yet been established. 
In this work, we develop, for the first time, a hierarchical semi-coherent search pipeline for XMRIs tailored to space-based gravitational-wave detectors, with a particular focus on the TianQin mission.
The pipeline combines a semi-coherent multi-harmonic $\mathcal{F}$-statistic with particle swarm optimization, and incorporates a novel eccentricity estimation method based on the relative power distribution among harmonics.
We validate the performance of the pipeline using simulated TianQin data for a Galactic Center XMRI composed of a brown dwarf and Sgr A*.
For a three-month observation, the pipeline successfully recovers the signal and achieves high-precision parameter estimation, including fractional uncertainties of $2.0\times10^{-6}$ in the orbital frequency, $2.9\times10^{-4}$ in the eccentricity, $2.5\times10^{-5}$ in the black-hole mass, and $5.6\times10^{-4}$ in the black-hole spin.
Our framework establishes a practical foundation for future XMRI searches with space-based detectors and highlights the potential of XMRIs as precision probes of stellar dynamics and strong-field gravity in the vicinity of supermassive black holes.
\end{abstract}

\maketitle

\acrodef{XMRI}{Extremely large mass-ratio inspiral}
\acrodef{EMRI}{Extreme mass-ratio inspiral}
\acrodef{BD}{brown dwarf}
\acrodef{SNR}{signal-to-noise ratio}
\acrodef{GW}{gravitational wave}
\acrodef{LISA}{Laser Interferometer Space Antenna}
\acrodef{TDI}{time-delay interferometry} 
\acrodef{PSD}{power spectral densities}
\acrodefplural{PSD}{power spectral densities}
\acrodef{EHT}{Event Horizon Telescope} 
\acrodef{PSO}{particle swarm optimization}
\acrodef{FF}{fitting factor}
\acrodef{GSF}{Gravitational Self-Force}
\acrodef{FEW}{FastEMRIWaveforms}
\acrodef{AK}{Analytic ``Kludge''}

\section{Introduction}

\acp{XMRI} consist of a \ac{BD} orbiting a supermassive black hole, such as Sgr~A*, on a tight relativistic trajectory \cite{Amaro-Seoane:2019umn,Gourgoulhon:2019iyu,Vazquez-Aceves:2022dgi,2025ApJ...987..208V,Seoane:2025hlg,Seoane:2025gfh,Fan:2025ymm}. 
Owing to the extreme mass ratio ($q \equiv M/m \sim 10^8$, where $M$ and $m$ denote the masses of Sgr A* and the \ac{BD}, respectively), the orbital evolution is exceptionally slow, allowing an \ac{XMRI} to complete up to $\sim 10^8$ cycles and remain observable for millions of years before merger. 
As a result, \acp{XMRI} emit long-lived, quasi-monochromatic \ac{GW} with frequencies in the millihertz band, well matched to the sensitivity of space-based \ac{GW} detectors \cite{Amaro-Seoane:2019umn,Gourgoulhon:2019iyu}. 
The clean dynamical nature of these systems makes \acp{XMRI} an ideal laboratory for precision tests of general relativity, and for measuring the mass and spin of the central black hole \cite{Fan:2025ymm,Amaro-Seoane:2020zbo}. 
Observations of multiple \acp{XMRI} can further probe the stellar environment around the central black hole and provide insights into stellar dynamics and evolution in galactic nuclei \cite{Gourgoulhon:2019iyu,Amaro-Seoane:2007osp,Eckart_stellar,2010RvMP...82.3121G,Genzel:2000mj}.

Theoretical studies based on stellar dynamics \cite{2013pss5.book..115K,2018A&A...609A..28B,2014CQGra..31x4007S,Bartko:2009qn,Preto:2009kd} in the Galactic Center suggest that the intrinsic formation rate of \acp{XMRI} is as low as $10^{-5} / {\rm yr}$.
However, owing to their exceptionally long inspiral timescales, the expected number of simultaneously observable \ac{XMRI} can reach the level of tens. 
In particular, it is predicted that around the Galactic centre, it may include around $\sim 15$ eccentric \acp{XMRI} and $\sim 5$ quasi-circular ones \cite{Amaro-Seoane:2019umn,2025ApJ...987..208V}.

Space-based \ac{GW} detectors enable direct observations in the millihertz frequency band by combining drag-free control of test masses with long-baseline laser interferometry. 
Several space missions have been proposed to explore this band, most notably the \ac{LISA} \cite{2017arXiv170200786A,LISA:2024hlh} and the TianQin \cite{TianQin:2015yph,TianQin:2020hid,Gong:2021gvw} mission, both of which target launch opportunities in the 2030s.
These detectors are expected to observe a wealth of \ac{GW} signals over years of operation \cite{2019PhRvD.100d3003W,2020PhRvD.102f3021H,2020PhRvD.101j3027L,2020PhRvD.102f3016F,2022PhRvD.105b2001L,2024PhRvD.109l4034Y,2023PhRvD.108h3023L,2024SCPMA..6779512C,2024PhRvD.110j3034Z}, creating a unique observational platform for studying compact-object dynamics in the strong-gravity regime \cite{LISA:2022yao,Li:2024rnk}.

Given the limited reach of space-based detectors, studies of \acp{XMRI} are predominantly focused on systems associated with the Milky Way's central black hole Sgr~A*. 
It is predicted that individual \ac{XMRI} signals can be detected with exceptionally high \ac{SNR}, typically ranging from $\mathcal{O}(10)$ up to $\mathcal{O}(10^{4})$, depending on the \ac{BD} mass and the orbital separation \cite{Amaro-Seoane:2019umn,Gourgoulhon:2019iyu}. 
These high signal strengths imply that \ac{XMRI} waveforms encode a wealth of information about the source parameters and the underlying strong-field dynamics. 
Indeed, theoretical forecasts suggest that key properties of the central massive black hole, including its mass and spin, can in principle be measured with extremely high precision\cite{Vazquez-Aceves:2022dgi,2025ApJ...987..208V}. 
Paradoxically, this very promise poses a fundamental challenge for data analysis: the combination of a high-dimensional parameter space and stringent accuracy requirements renders traditional template-based searches an impractical ``needle-in-a-haystack'' problem. 
Without dedicated data-analysis methodologies, the theoretical precision promised by \acp{XMRI} remains unattainable in practice.

To address these complexities, significant efforts have been devoted to modeling \ac{EMRI} waveforms and developing robust search algorithms. Waveform modeling has evolved from computationally efficient \ac{AK} templates \cite{Babak:2006uv,Chua:2017ujo} to high-fidelity relativistic frameworks based on the gravitational self-force program \cite{vandeMeent:2017bcc,Pound:2019lzj,Wardell:2021fyy,Katz:2021yft}. On the data-analysis front, a diverse array of strategies has been explored, including Markov chain Monte Carlo methods \cite{Babak:2009ua,MockLISADataChallengeTaskForce:2009wir}, time-frequency track searches \cite{Gair:2007bz,Speri:2025ucn}, phenomenological template searches \cite{Wang:2012xh} and more recently, hierarchical searches \cite{2024PhRvD.109l4034Y} and deep-learning-based architectures \cite{Zhang:2022xuq,Yun:2023vwa}. Despite these extensive efforts, the high dimensionality and complexity of \ac{EMRI} waveforms impose stringent demands on data analysis methodologies, and a universally satisfactory solution has yet to be established. While these advancements have laid the foundation for \ac{EMRI} science, the specialized requirements of \ac{XMRI} searches—characterized by even more extreme mass ratios and year-long quasi-monochromatic signals—motivate us to develop a computationally efficient search strategy by referring to mature methodologies established for other quasi-monochromatic signals.

The data-analysis strategies for quasi-monochromatic \ac{XMRI} signals draw heavily from other quasi-monochromatic \ac{GW} search methodologies. Historically, the $\mathcal{F}$-statistic was developed for ground-based pulsar searches to analytically account for Doppler modulations \cite{Jaranowski:1998qm}, serving as a cornerstone alongside other techniques such as the Hough transform and PowerFlux pipelines \cite{Arnaud:2007vr,MockLISADataChallengeTaskForce:2007iof,MockLISADataChallengeTaskForce:2009wir}. This framework has since been successfully adapted for space-based missions to resolve the Galactic population of double white dwarfs \cite{Blaut:2009si,Zhang:2021htc,Lu:2022ywf}. To manage the immense computational cost and ensure robustness over year-long observations, semi-coherent methods are typically employed, where data are partitioned into shorter segments to balance sensitivity with search efficiency \cite{Dreissigacker:2018afk,Mirasola:2024lcq}. Consequently, these established methodologies for quasi-monochromatic \ac{GW} searches provide a robust foundation that can be effectively leveraged for the data-analysis requirements of \acp{XMRI}.

At present, the mass and spin of Sgr~A* are constrained through a combination of electromagnetic observations \cite{Daly:2023axh}, most notably stellar dynamics \cite{Ghez:2008ms,Gillessen:2008qv,Fragione:2020khu} and horizon-scale imaging by \ac{EHT} \cite{EventHorizonTelescope:2022wkp,EventHorizonTelescope:2022apq,EventHorizonTelescope:2022urf,EventHorizonTelescope:2022wok,EventHorizonTelescope:2022exc,EventHorizonTelescope:2024hpu}. 
While these approaches have delivered remarkable insights into the nature of Sgr~A*, their ability to precisely determine black hole spin and to break parameter degeneracies remains fundamentally limited by astrophysical systematics and modelling uncertainties \cite{Ciurlo:2025qns}. 
In contrast, \acp{GW} emitted by \acp{XMRI} offer a direct probe of the spacetime geometry, thereby offering the potential to place unprecedented constraints on the properties of Sgr~A*, enabling precision tests of strong gravity that are inaccessible to existing electromagnetic techniques. Furthermore, the dynamical evolution of such \ac{XMRI} systems is potentially linked to the frequent electromagnetic flares observed at the Galactic Center, suggesting a promising synergy for multi-messenger studies that combine \ac{GW} signals with high-cadence electromagnetic monitoring of Sgr~A* \cite{Seoane:2025hlg}. Such multi-messenger prospects further underscore the necessity of a robust search pipeline capable of processing long-duration \ac{XMRI} signals from space-based observatories.

In this work, we present the first development of a hierarchical semi-coherent search pipeline of anticipated \ac{XMRI} signals, assuming the TianQin detector.
The method combines $\mathcal{F}$-statistic, \ac{PSO} search algorithm, with a novel multi-harmonic eccentricity estimation method.
This paper is organized as follows. 
In Section~\ref{sec:theoretical_modeling}, we describe the theoretical modeling of \ac{XMRI} systems, including the waveform and the relevant properties of space-based gravitational-wave detectors, as well as their noise characteristics. 
Section~\ref{sec:method} introduces the data-analysis methodology, covering the statistical basics, the fully-coherent and semi-coherent $\mathcal{F}$-statistic, the eccentricity estimation strategy, and the \ac{PSO} algorithm adopted in the pipeline. 
Section~\ref{sec:implementation} presents the implementation of the search pipeline, including data simulation, template construction, and the hierarchical semi-coherent search procedure. 
The results of our analysis are presented in Section~\ref{sec:results}, and we discuss the implications of our findings in Section~\ref{sec:discussion}. 
Finally, we summarize our conclusions in Section~\ref{sec:conclusion}.

\section{Theoretical modeling}\label{sec:theoretical_modeling}

In this section, we aim to establish the signal and noise model adopted throughout this work for the analysis of \ac{XMRI} signals in space-based gravitational-wave detectors.
We first describe the gravitational-wave waveform of \acp{XMRI}, then introduce the detector and \ac{TDI} observables, and finally specify the instrumental noise model used in our data analysis.

\subsection{XMRI waveform model}\label{subsec:XMRI_model}

\begin{figure}[htbp]
    \centering
    \includegraphics[width=1.0\linewidth]{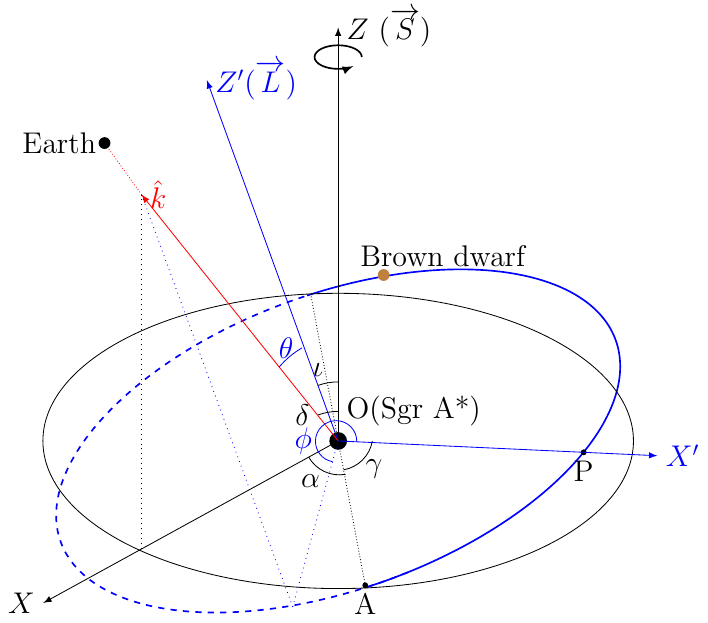}
    \caption{Geometry of an eccentric \ac{XMRI} system in the black-hole spin–aligned frame $(X,Y,Z)$ and the orbital frame $(X',Y',Z')$. $Z$-axis is parallel to the spin of Sgr~A* ($\vec{S}$), while $Z'$-axis aligned with the orbital angular momentum $\vec{L}$, where the $X'$-axis points towards the pericenter $P$. The misalignment between $\vec{L}$ and $\vec{S}$ is characterized by the inclination angle $\iota$. The unit vector $\hat{k}$ denotes the propagation direction of the \acp{GW} toward Earth, with $(\theta, \phi)$ being its corresponding polar and azimuthal angles in the orbital frame, and $\delta$ being its angle relative to $\vec{S}$. The geometry is further specified by the longitude of the ascending node (point $A$)  $\alpha$ and the argument of pericenter $\gamma$.}
    \label{fig:coordinate}
\end{figure}

Throughout this manuscript, we consider an \ac{XMRI} system that consists of a brown dwarf inspiraling around the supermassive black hole Sgr~A* at the Galactic Center. 
We illustrate the configurations in Fig.~\ref{fig:coordinate}. 
To characterize the geometry of the system, we introduce two reference frames: a black-hole frame $(X,Y,Z)$  (in black) aligned with the spin of Sgr~A*, and an orbital frame $(X',Y',Z')$ (in blue) attached to the instantaneous orbital plane of the inspiraling \ac{BD}. 
For the sake of generality, we allow the orbital angular momentum $\vec{L}$ of the \ac{BD} to be misaligned with the massive black-hole spin $\vec{S}$~\cite{2025ApJ...987..208V}.
The relative orientation between these two orbits is described by the inclination angle $\iota$ between $\vec{L}$ and $\vec{S}$, while the direction of \ac{GW} propagation toward the Earth, denoted by the unit vector $\hat{k}$, forms an angle $\delta$ with respect to the black-hole spin. 
We also express $\hat{k}$ in the orbital frame by the polar angle $\theta$ and the azimuth angle $\phi$. 
We note that our definition of $\phi$ is related to the angle $\phi_M$ used by Maggiore \cite{Maggiore:2007ulw} via $\phi = \pi/2 - \phi_M$. Consequently, the trigonometric functions in Eq.~\eqref{xmri_sig} differ by a phase shift. Furthermore, our polarization basis is rotated relative to that in Ref.~\cite{Maggiore:2007ulw} by a polarization angle of $\pi/2$, which accounts for the overall sign difference in both $h_{n,+}$ and $h_{n,\times}$.

The orbital configuration is further specified by $\alpha$, or the longitude of the ascending node $A$, and $\gamma$, or the argument of pericenter $P$. 
$\alpha$ and $\gamma$ encode the precession of the orbital plane and the argument of pericenter, respectively.

The orbital dynamics of an inspiraling \ac{BD} can be primarily characterized by two parameters: the orbital frequency $f_0$ and the eccentricity $e$. 
Under the assumption that the motion is dominated by the gravitational field of the central massive black hole, these quantities are related to the orbital geometry through Kepler's third law \cite{Maggiore:2007ulw,Gourgoulhon:2019iyu}
\begin{equation}
    f_0 = \frac{1}{2\pi} \sqrt{\frac{GM}{r_a^3}},
\end{equation}
where $G$ is the gravitational constant and $r_a$ denotes the semi-major axis. 
For an eccentric orbit, the geometry is further specified by the semi-minor axis $r_b$ and the periapsis distance $r_p$
\begin{equation}
    r_b = r_a\sqrt{1-e^2}, \quad r_p = r_a(1-e).
\end{equation}

The presence of orbital eccentricity leads to a rich multi-harmonic structure in the emitted \ac{GW} signal \cite{PhysRev.131.435}. 
In the orbital frame, the \ac{GW} waveform from an eccentric \ac{XMRI} can therefore be expressed as a superposition of harmonics of the orbital frequency \cite{Barack:2003fp}. 
For a given harmonic index $n$, the plus and cross polarizations take the form \cite{Maggiore:2007ulw,1995MNRAS.274..115M,Barack:2003fp,PhysRev.131.435}
\begin{equation}
\begin{aligned}
&h_{n,+}(t;\theta,\phi) \\ & =\frac{{\mu}\omega_n^2}{d_\mathrm{L}} \frac{G}{c^4} 
\biggl[ A_n(\sin^2\phi - \cos^2\phi \cos^2\theta) \cos\Phi_{n}(t) \\
& + B_n(\cos^2\phi - \sin^2\phi \cos^2\theta) \cos\Phi_{n}(t) \\
& - C_n \sin 2\phi (1 + \cos^2\theta) \sin\Phi_{n}(t) \biggr] ,\\
&h_{n,\times}(t;\theta,\phi) \\&=  \frac{{\mu}\omega_n^2}{d_\mathrm{L}} \frac{G}{c^4} 
\biggl[ (A_n - B_n) \sin 2\phi \cos\theta \cos\Phi_{n}(t) \\
& - 2C_n \cos 2\phi \cos\theta \sin\Phi_{n}(t) \biggr].
\end{aligned}
\label{xmri_sig}
\end{equation}
Here $c$ denotes the speed of light, $\omega_{n,0} = 2\pi n f_0$ denotes the angular frequency of the $n$-th harmonic, and the phase evolution is given by
$\Phi_n(t) = \omega_{n,0} t + \frac{1}{2}\dot{\omega}_n t^2 + n\Phi_0 + \mathcal{O}(\ddot{\omega_n} t^3)$,
where $\dot{\omega}_n$ ($\ddot{\omega_n}$) represents the first (second) order time derivative of the harmonic angular frequency and $\Phi_0$ is the initial orbital phase.
For Sgr~A*, we adopt a luminosity distance of $d_\mathrm{L} = 8~\mathrm{kpc}$ \cite{2019A&A...625L..10G,2019ApJ...885..131R}.
The reduced mass of the system is defined as
\begin{equation}
\mu = \frac{Mm}{M+m} \approx m,
\label{reduced_mass}
\end{equation}
where we adopt $M=4\times10^6~\mathrm{M_{\odot}}$ \cite{Ghez:2008ms,GRAVITY:2018ofz} and $m=0.05~\mathrm{M_{\odot}}$~\cite{Vazquez-Aceves:2022dgi,2025ApJ...987..208V} corresponding to the masses of Sgr~A* and the inspiraling \ac{BD}, respectively.

The coefficients $A_n$, $B_n$, and $C_n$ encode the dependence of the waveform on the orbital eccentricity $e$ and determine how the \ac{GW} power is distributed among different harmonics.
They are expressed in terms of Bessel functions as \cite{Maggiore:2007ulw}
\begin{equation}
\begin{aligned}
A_n = &\frac{r_a^2}{n}[ J_{n-2}(n e) - J_{n+2}(n e) - 2e J_{n-1}(n e) \\
&+ 2e J_{n+1}(n e)], \\
B_n = &\frac{r_b^2}{n} [ J_{n+2}(n e) - J_{n-2}(n e) ], \\
C_n = &\frac{r_ar_b}{n} [ J_{n-2}(n e) + J_{n+2}(n e) - e J_{n-1}(n e)\\
&- e J_{n+1}(n e) ].
\label{bessel}
\end{aligned}
\end{equation}

The orbital parameters of the \ac{XMRI} system, including the orbital frequency $f_0$ and eccentricity $e$, are not constant over time as the \ac{BD} inspirals toward Sgr~A*. 
These parameters are subject to secular evolution due to the combined effects of gravitational radiation reaction and the spin of the central black hole $s$. 
Specifically, the orbital frequency $f_0(t)$ and eccentricity $e(t)$ change as a result of \ac{GW} emission \cite{osti_4688457}, while the orientation of the orbit, characterized by the longitude of the ascending node $\alpha(t)$ and the argument of pericenter $\gamma(t)$, is affected by relativistic precession induced by the black hole's spin \cite{Barack:2003fp}. 
\begin{subequations}
\label{eq:full_orbital_dynamics}
\begin{align}
    \frac{\mathrm{d} f_0}{\mathrm{d} t} = & \, \frac{48}{5\pi} \left( \frac{G \mathcal{M}_c}{c^3} \right)^{5/3} (2\pi f_0)^{11/3} (1-e^2)^{-7/2} \nonumber \\
    & \times \left( 1 + \frac{73}{24}e^2 + \frac{37}{96}e^4 \right) , \label{eq:df0dt} \\[10pt]
    \frac{\mathrm{d} e}{\mathrm{d} t} = & \, - \frac{304}{15} e \left( \frac{G \mathcal{M}_c}{c^3} \right)^{5/3} (1-e^2)^{-5/2} (2\pi f_0)^{8/3} \nonumber \\
    & \times \left( 1 + \frac{121}{304}e^2 \right) , \label{eq:dedt} \\[10pt]
    \frac{\mathrm{d} \gamma}{\mathrm{d} t} = & \, 6 \pi f_0 \left( \frac{2 \pi G M f_0}{c^3} \right)^{2/3} (1-e^2)^{-1} \nonumber \\
    & \times \left[ 1 + \frac{1}{4} \left( \frac{2 \pi G M f_0}{c^3} \right)^{2/3} \frac{26 - 15 e^2}{1 - e^2} \right] \nonumber \\
    & - 12 \pi f_0 s \cos \iota \left( \frac{2 \pi G M f_0}{c^3} \right) (1-e^2)^{-3/2} , \label{eq:dgammadt} \\[10pt]
    \frac{\mathrm{d} \alpha}{\mathrm{d} t} = & \, 4 \pi f_0 s \left( \frac{2 \pi G M f_0}{c^3} \right) (1-e^2)^{-3/2}, \label{eq:dalphadt}
\end{align}
\end{subequations}
where $\mathcal{M}_c = (Mm)^{3/5}/(M+m)^{1/5}$ denotes the chirp mass. 

Given the slow evolution of \ac{XMRI} systems, the time dependence of the orbital parameters can be well described by a linear approximation\cite{Lin2022Identifying},
\begin{equation}
x(t) = x_0 + \dot{x}\, t, \quad x \in \{ f_0, e, \alpha, \gamma \}.
\end{equation}
This approximation enables an efficient construction of search templates without compromising the level of accuracy required in this work.

For \acp{XMRI} around Sgr~A*, the extreme mass ratio ($q \sim 10^8$) ensures that the system resides deep within the adiabatic regime, evolves very slowly, and remains close to geodesic motion over the duration of a space-based mission.
Consequently, the \ac{AK} waveform is an appropriate choice for this study, as it offers a physically grounded and computationally tractable framework for modeling such long-lived signals.
While high-fidelity models such as \ac{GSF} \cite{vandeMeent:2017bcc,Wardell:2021fyy} and the \ac{FEW} framework \cite{Katz:2021yft,Khalvati:2024tzz} offer superior relativistic fidelity, they currently face limitations in computational tractability and parameter space coverage of the eccentric and inclined parameter space investigated here.
Importantly, our search pipeline is designed to be modular; should more precise waveform models become available in the future, they can be directly integrated as plug-and-play replacements without altering the underlying search strategy.

\subsection{Space-based gravitational-wave detectors}

\ac{XMRI} systems emit long-lived \ac{GW} signals predominantly in the millihertz frequency band, which can be detected by space-based \ac{GW} observatories \cite{Amaro-Seoane:2019umn,Amaro-Seoane:2020zbo}. 
Current and planned missions targeting this frequency range include the geocentric TianQin mission \cite{TianQin:2015yph,TianQin:2020hid,Li:2024rnk} and the heliocentric LISA \cite{2017arXiv170200786A,LISA:2022yao}. 
In this work, we focus on TianQin as a representative example and develop an \ac{XMRI} data-analysis pipeline tailored to its detector configuration and observational strategy.

TianQin is a space-based gravitational-wave detector consisting of three satellites arranged in a regular-triangular constellation on a geocentric orbit with an orbital altitude of $\sim 10^{5}\,\mathrm{km}$ and an arm length of $L=\sqrt{3}\times10^{8}\,\mathrm{m}$. 
It is designed to be sensitive in the frequency band from $0.1\,\mathrm{mHz}$ to $1\,\mathrm{Hz}$. 
The detector configuration adopts a geocentric orbit in conjunction with a fixed orientation of the detector plane relative to the reference source J0806 (also known as HM Cancri), an ultra-compact verification binary that serves as a calibration source for space-based \ac{GW} detectors due to its precisely known sky location and orbital period \cite{Stroeer:2006rx}. 
Moreover, TianQin adopts a ``3+3'' operational scheme, alternating between three months of scientific observation and three months of maintenance to ensure thermal and orbital stability \cite{TianQin:2015yph,TianQin:2020hid}. 
The nominal mission lifetime is $5$ years, corresponding to an effective observation time of approximately $2.5$ years.

\subsection{Time delay interferometry and noise model}

For space-based \ac{GW} detectors, the dominant contribution to the raw phasemeter measurements arises from intrinsic laser-frequency fluctuations \cite{Prince:2002hp,Cornish:2002rt}. 
These laser frequency noises are many orders of magnitude larger than the target \ac{GW} signals.
\ac{TDI} suppresses this noise by forming linear combinations of data streams with appropriate light-travel-time delays. This procedure effectively constructs a virtual equal-arm interferometer, in which the laser frequency noise from different arms cancels out  \cite{1999ApJ...527..814A,Tinto:2020fcc}.

In this work, we adopt the first-generation \ac{TDI} observables, which assume quasi-static armlengths. The explicit construction of the corresponding \ac{TDI} link combinations can be found in Refs.~\cite{Tinto:2020fcc,Li:2023szq} and is not repeated here.
This approximation is adequate for our purposes, since our analysis focuses on the statistical properties of the detection statistic, and higher-generation \ac{TDI} effects are not expected to qualitatively affect the conclusions.
In the first-generation \ac{TDI} framework, the unequal-arm Michelson-type observables are commonly denoted as $X$, $Y$, and $Z$, which are constructed from appropriately time-delayed one-way inter-satellite laser links \cite{Tinto:2020fcc,1999ApJ...527..814A,Wang:2025ejp,Li:2023szq}.

By assuming that the instrumental noise in each interferometer arm has identical statistical properties and is uncorrelated between different arms, these Michelson-like observables can be further combined into noise-orthogonal channels
\begin{equation}
\begin{aligned}
A &= \frac{Z-X}{\sqrt{2}}, \\
E &= \frac{X - 2Y+Z}{\sqrt{6}}, \\
T &= \frac{X + Y + Z}{\sqrt{3}},
\end{aligned}
\end{equation}
which are convenient for sensitivity estimates and data analysis because the $A$ and $E$ channels capture most of the signal power in the mHz band while the $T$ channel is largely insensitive to \ac{GW} signals \cite{Tinto:2020fcc,Li:2023szq}.
In the following analysis, for the sake of convenience, we focus on the $A$ and $E$ channels and ignore the $T$ channel.

The residual instrumental noise in space-based detectors is primarily attributed to two sources: test-mass acceleration noise that dominates at low frequencies, and interferometer position or displacement noise that becomes more significant at higher frequencies \cite{TianQin:2015yph}. 
We assume the noise to be stationary and Gaussian \cite{1999ApJ...527..814A,Wang:2025ejp,Tinto:2020fcc}.
In this work, we adopt the \acp{PSD} for the noise-orthogonal \ac{TDI} channels $(A,E,T)$ defined by \texttt{GWSpace} \cite{Li:2023szq}, and generate the corresponding noise realizations for our analysis. For completeness, the explicit expressions of the TianQin noise PSDs are summarized in Ref.~\cite{Li:2023szq} and are not repeated here.

\section{Methodology}\label{sec:method}

In this section, we describe the methodology used in this work, which includes the fully-coherent and semi-coherent $\mathcal{F}$-statistic \cite{Jaranowski:1998qm,Cornish:2003vj,Cutler:2005hc,Prix2009,Prix:2009tq,Covas:2022xyd,Fu:2024cpu,2024PhRvD.110j3026B}, and the \ac{PSO} technique applied to the optimization problem \cite{4223164,mohanty2018swarm}. 
The \ac{PSO} method could encounter difficulty in optimizing certain parameters like the orbital eccentricity $e$.
Therefore, we also develop a method to determine the eccentricity of the \acp{XMRI} based on the relative powers of different harmonics.

\subsection{Statistical basics}

The detector output can be modeled as
\begin{equation}
x(t) = h(t) + n(t),
\end{equation}
where $h(t)$ is a possible \ac{GW} signal and $n(t)$ is assumed to be a zero-mean, stationary Gaussian noise process.

Under the null hypothesis $H_0$ (noise only), the probability density of observing $x(t)$ is \cite{Prix2009}
\begin{equation}\label{eq:H0}
P(x \mid H_0) \propto \exp\Big[-\frac{1}{2}\langle x \mid x \rangle\Big],
\end{equation}
while under the alternative hypothesis $H_1$ (signal plus noise), it is
\begin{equation}
P(x \mid H_1) \propto \exp\Big[-\frac{1}{2}\langle x-h \mid x-h \rangle \Big].
\end{equation}

The likelihood ratio, defined as
\begin{equation}
\Lambda(x;h) \equiv \frac{P(x \mid H_1)}{P(x \mid H_0)} = \exp\Big[\langle x \mid h \rangle - \frac{1}{2}\langle h \mid h\rangle\Big],
\end{equation}
provides the optimal test statistic according to the Neyman-Pearson lemma.
This leads to the standard expression for the log-likelihood ratio used in matched filtering\cite{Finn:1992wt}
\begin{equation}\label{eq:log_likelihood_ratio}
\ln \Lambda(x;h) = \langle x \mid h \rangle - \frac{1}{2}\langle h \mid h\rangle,
\end{equation}
where the inner product is defined by
\begin{equation}\label{eq:inner_product}
\langle a \mid b\rangle \equiv 4 \,\mathrm{Re} \int_0^{+\infty} \frac{\tilde{a}(f) \tilde{b}^*(f)}{S_n(f)} \, {\rm d}f.
\end{equation}
Here, $\tilde{a}(f)$ and $\tilde{b}(f)$ denote the Fourier transforms of
$a(t)$ and $b(t)$; $S_n(f)$ is the one-sided \ac{PSD};
$^*$ denotes complex conjugation; and $\mathrm{Re}$ indicates the real part. 
This noise-weighted inner product induces a natural norm on the signal space,
which allows one to define the \ac{SNR} of a \ac{GW} signal.

For a signal $h(t)$, the \emph{optimal} SNR \cite{Cutler:1994ys} is given by
\begin{equation}
\rho \equiv \sqrt{\langle h \mid h \rangle} .
\end{equation}

The inner product further allows us to define the \ac{FF} \cite{Owen:1995tm}, which quantifies the maximum similarity between a signal $h(\theta)$ and a template $h(\theta^{\prime})$
\begin{equation}
\text{FF}(\theta) \equiv \max_{\theta^{\prime}} \frac{\langle h(\theta) \mid h(\theta^{\prime}) \rangle}{\sqrt{\langle h(\theta) \mid h(\theta) \rangle \langle h(\theta^{\prime}) \mid h(\theta^{\prime}) \rangle}}.
\end{equation}
The \ac{FF} provides a quantitative measure of the template's fidelity. 
It represents the fractional loss in the recoverable \ac{SNR} due to model inaccuracies or parameter offsets. 

\subsection{\texorpdfstring{Fully-coherent and semi-coherent $\mathcal{F}$}{F}-statistic} \label{subsec:F-statistic}

For \ac{XMRI} signals, where each harmonic can be treated as quasi-monochromatic, the \ac{GW} power is concentrated in a narrow frequency band around the $n-$th harmonic frequency $f_{0,n}$. 
In this narrow band, the noise \ac{PSD} can be approximated as constant, $S_n(f) \approx S_n(f_{0,n})$. 
According to Parseval's theorem, the inner product then reduces to
\begin{equation}
\langle a \mid b \rangle \approx \frac{2}{S_n(f_{0,n})} \int_0^{T_{\rm obs}} a(t) b(t) \, {\rm d}t,
\end{equation}
where $T_{\rm obs}$ is the observation time of the source.

Since the harmonics of \ac{XMRI} signals are quasi-monochromatic, it is convenient to rewrite the signal with detector response so that the extrinsic and intrinsic parameters can be explicitly separated.
In particular, the strain contribution from the $n$-th harmonic can be expressed as
\begin{equation}\label{eq:factored_waveform}
h_n(t;\mathcal{A}_n,\lambda) = \sum^2_{\mu=1} \mathcal{A}_n^{\mu} h_{\mu,n}(t;\lambda),
\end{equation}
where $\mathcal{A}_n^{\mu}$ collects the extrinsic amplitude parameters, while $h_{\mu,n}(t;\lambda)$ depends only on the intrinsic parameters $\lambda$. Explicit expressions for these extrinsic amplitudes and the corresponding intrinsic basis functions are detailed in Appendix~\ref{app:extrinsic}.

Inserting the factored expression Eq.~\eqref{eq:factored_waveform} into Eq.~\eqref{eq:log_likelihood_ratio} gives the log-likelihood for the $n-$th harmonic,
\begin{equation}
    \ln \Lambda_n(x; \mathcal{A}_n, \lambda)
= \mathcal{A}^{\mu}_n x_{\mu,n}
- \tfrac{1}{2} \mathcal{A}^{\mu}_n \mathcal{A}^{\nu}_n \mathcal{M}_{\mu\nu,n} \, ,
\label{eq:log_likelihood_new}
\end{equation}
with implicit summation over \(\mu,\nu=1,2\), where we have defined
\begin{equation}
    x_{\mu,n} \equiv \langle x \mid h_{\mu,n}\rangle, \qquad 
    \mathcal{M}_{\mu\nu,n} \equiv \langle h_{\mu,n}\mid h_{\nu,n}\rangle .
\end{equation}

Maximizing \(\ln\Lambda_n\) with respect to the extrinsic amplitude parameters \(\mathcal{A}_n^{\mu}\) leads to 
\begin{equation}
\frac{\partial \ln \Lambda_n}{\partial \mathcal{A}_n^{\mu}} = 0
\quad \Rightarrow \quad
\mathcal{A}_{\mathrm{ML},n}^{\mu}
= \mathcal{M}_n^{\mu\nu} \, x_{\nu,n},
\label{eq:ml_extrinsic_estimator}
\end{equation}
where \(\mathcal{M}_n^{\mu\alpha}\mathcal{M}_{n,\alpha\nu}=\delta^\mu{}_\nu\). Here, $\mathcal{A}_{\mathrm{ML},n}^{\mu}$ represent the maximum-likelihood estimators for the $n$-th harmonic \cite{Prix2009}. Substituting \(\mathcal{A}_{\mathrm{ML},n}^{\mu}\) back into Eq.~\eqref{eq:log_likelihood_new} yields the detection statistic (the $\mathcal{F}$-statistic) on a single harmonic 
\begin{equation}
    2\mathcal{F}_n(x;\lambda) \equiv 2\ln \Lambda_n(x; \mathcal{A}_{\mathrm{ML},n}, \lambda)
= x_{\mu,n}\,\mathcal{M}_n^{\mu\nu}\,x_{\nu,n}.
\label{eq:F_stat_n}
\end{equation}

In order to obtain the overall detection statistic, one needs to combine contributions from all harmonics and all TDI channels.
For the three noise-orthogonal TDI channels $(A, E, T)$, both $A/E$ signal-sensitive channels can independently contribute to the likelihood ratio.
Therefore, the corresponding inner products are additive. 
For each harmonic \(n\) we define
\begin{eqnarray}
    x_{AE,\mu;n} & =& x_{A,\mu;n} + x_{E,\mu;n}, \\ 
    \mathcal{M}_{AE,\mu\nu;n} &=& \mathcal{M}_{A,\mu\nu;n} + \mathcal{M}_{E,\mu\nu;n},
\end{eqnarray}
and the combined $AE$ statistic for harmonic \(n\) reads \cite{Cutler:2005hc}
\begin{equation}
    2\mathcal{F}_{AE,n} = x_{AE,\mu;n}\,\mathcal{M}_{AE;n}^{\mu\nu}\,x_{AE,\nu;n}.
\end{equation}

Similarly, as different harmonics contribute to the detection statistic independently, the total coherent $\mathcal{F}$-statistic for a given segment is obtained by summing over all detectable harmonics $n \in [n_{\min}, n_{\max}]$:
\begin{equation}
    2\mathcal{F}_{AE,\mathrm{tot}} = \sum_{n=n_{\min}}^{n_{\max}} 2\mathcal{F}_{AE,n},
\end{equation}
where $n_{\max} = n_{\min} + N - 1$, and $N$ denotes the total number of harmonics included in the calculation. The specific criteria for choosing $n_{\min}$ and $n_{\max}$ are discussed in Sec.~\ref{subsec:pipeline}.

In principle, one can apply the above statistic to the entire data fully coherently.
However, the hardware cost of the fully coherent search becomes computationally prohibitive for large parameter spaces and narrow-template-bank mismatch requirements.
Semi-coherent methods mitigate this by dividing the full dataset into \(K\) shorter coherent segments, computing the coherent \(\mathcal{F}\)-statistic in each segment, and then incoherently combining these segment-wise statistics. 
The semi-coherent approach broadens the effective search peaks in parameter space (reducing sensitivity to template mismatch) while retaining much of the coherent sensitivity gain; in this work, we form the semi-coherent statistic simply as the sum of segment-wise \(2\mathcal{F}_{\, AE,\mathrm{tot}}\) values over all segments,
\begin{equation}
    2 \mathcal{F}_{\, AE}^{\,\mathrm{semi}}
    = \sum_{k=1}^{K} 2 \mathcal{F}_{\, AE\mathrm{, tot}}^{(k)}
    = \sum_{k=1}^{K} \sum_{n=n_{\min}}^{n_{\max}} 2 \mathcal{F}_{AE,n}^{(k)}.
\end{equation}

Under the assumption of Gaussian noise, the semi-coherent statistic $2\mathcal{F}_{\mathrm{AE}}^{\mathrm{semi}}$ follows a $\chi^2$ distribution.
In the absence of a signal, it is distributed as a central $\chi^2$, whereas in the presence of a perfectly matched signal, it follows a non-central $\chi^2$ distribution.
The number of degrees of freedom is determined by the fact that the $A$ and $E$ TDI channels are combined at the likelihood level and share a single set of extrinsic amplitude parameters, contributing two amplitude degrees of freedom per harmonic and per semi-coherent segment.
Summing over $N$ harmonics and $K$ independent semi-coherent segments, therefore, yields a total of $2KN$ degrees of freedom.
The expectation value and variance of the semi-coherent statistic are then given by
\begin{subequations}
\label{eq:mean_var_2F_set}
\begin{align}
    \left\langle 2\mathcal{F}_{AE}^{\mathrm{semi}} \right\rangle
    &= 2KN + \rho_{\mathrm{tot}}^2, \label{eq:mean_2F} \\[8pt]
    \mathrm{Var}\!\left(2\mathcal{F}_{AE}^{\mathrm{semi}}\right)
    &= 4KN + 4\rho_{\mathrm{tot}}^2, \label{eq:var_2F}
\end{align}
\end{subequations}
where $\rho_{\mathrm{tot}}^2$ denotes the total non-centrality parameter, defined as the sum of the squared signal-to-noise ratios over all harmonics and all semi-coherent segments.
The special case $K=1$ corresponds to the fully coherent $\mathcal{F}$-statistic.

Before proceeding further, it is useful to examine the behavior of the $\mathcal{F}$-statistic in a controlled setting in order to validate its mathematical construction.
Such a test provides a direct consistency check of the analytical derivation, independent of any specific search strategy or optimization method.
A particularly simple and transparent limit is the noise-free case, where the matched-filtering formalism predicts an exact relation between the $\mathcal{F}$-statistic and the optimal signal-to-noise ratio.

\begin{figure}[htbp]
    \centering
    \includegraphics[width=1.0\linewidth]{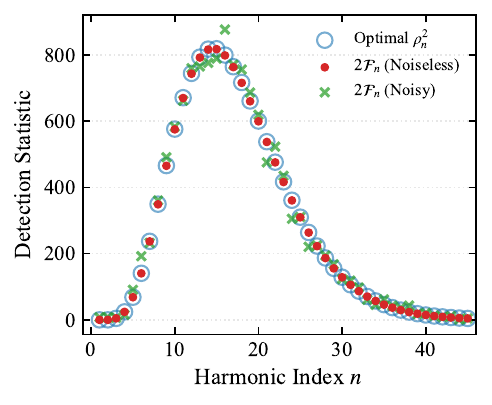}
    \caption{The detection statistic $2\mathcal{F}_n$ versus harmonic index $n$. 
    Noiseless results (red dots) coincide with the optimal \ac{SNR} squared $\rho_n^2$ (blue open circles), while noisy realizations are shown as green crosses.}
    \label{fig:F_SNR}
\end{figure}

In the absence of noise, the $\mathcal{F}$-statistic for the $n$th harmonic reduces to the squared signal-to-noise ratio, $2\mathcal{F}_n = \rho_n^2$, as implied by its statistical construction.
Fig.~\ref{fig:F_SNR} compares the numerically evaluated $2\mathcal{F}_n$ (red dots) with the corresponding theoretical values $\rho_n^2$ (blue open circles) across all harmonics.
The exact agreement observed in the noiseless case confirms both the internal consistency of the $\mathcal{F}$-statistic formalism and the correctness of its numerical implementation.
In practical applications, the statistic is evaluated on data containing instrumental noise, yielding a single realization of the underlying random variable.

\subsection{Multi-harmonic eccentricity estimation}\label{sec:eccentricty_estimation}

In principle, one can count on the $\mathcal{F}$-statistic-based searches to identify the best-matched signal parameters. 
In practice, however, we find that for certain parameters—most notably the orbital eccentricity $e$—direct optimization using the $\mathcal{F}$-statistic alone can be inefficient and prone to large biases.
In this work, we propose a novel method to estimate $e$ through the comparing relative powers among different harmonics, which we elaborate in this subsection.

Notice that the eccentricity enters the waveform primarily through the relative distribution of signal power among different harmonics. 
On the other hand, the signal-to-noise ratio is deeply linked to the $\mathcal{F}$-statistic value.
As illustrated in Fig.~\ref{fig:F_SNR}, while the $2\mathcal{F}_n$ values in the presence of noise (green crosses) inevitably deviate from the optimal $\rho_n^2$ (blue open circles), their overall distribution across the harmonic index $n$ remains consistent.
This structural similarity ensures that the harmonic power profile, captured by the $\mathcal{F}$-statistic, provides a robust signature for eccentricity estimation even in noisy data.
Therefore, we utilize the relative values of $2\mathcal{F}_n$ measured across different harmonics as a probe of the underlying $e$.
As discussed in Section \ref{subsec:XMRI_model}, for an eccentric compact object inspiraling into a massive black hole, the \ac{GW} signal can be decomposed into harmonics of the orbital frequency.
Under the stationary-phase approximation, the squared \ac{SNR} accumulated by the $n-$th harmonic can be written as \cite{Wagg:2021sgn}
\begin{equation}
\rho_n^2 = \frac{1}{4} \frac{h_n^2 T_{\rm obs}}{S_n(f_n)} ,
\label{eq:snr_sum}
\end{equation}
where $S_n(f_n)$ is evaluated at the harmonic frequency $f_n = n f_0$.  
The strain amplitude of the $n$-th harmonic $h_n$ is given by \cite{Wagg:2021sgn,Finn:2000sy}
\begin{equation}
h_n^2
=
\frac{2^{28/3}}{5}
\frac{(G\mathcal{M}_c)^{10/3}}{c^8 d_{\mathrm{L}}^2}
\frac{g(n,e)}{n^2}
(\pi f_0)^{4/3},
\label{eq:hn}
\end{equation}
where $g(n,e)$ \cite{PhysRev.131.435} encodes the eccentricity dependence of the harmonic power.
Combining Eqs.~(\ref{eq:snr_sum}) and (\ref{eq:hn}), the expected \ac{SNR} contribution of each harmonic can be expressed as
\begin{equation}
\rho_n^2
=
\frac{2^{28/3}}{20}
\frac{(G\mathcal{M}_c)^{10/3} T_{\rm obs}}{c^8 d_{\mathrm{L}}^2}
(\pi f_0)^{4/3}
\frac{g(n,e)}{n^2 S_n(n f_0)} .
\label{eq:snr_n}
\end{equation}

We notice that for different harmonics, the variation of the $\rho_n^2$ is entirely determined by the harmonic index $n$ and the orbital eccentricity $e$. 
All remaining parameters, including the chirp mass, luminosity distance, and observation time, enter only as an overall multiplicative factor common to all harmonics. 
We therefore rewrite the expected SNR contribution of the $n$-th harmonic as
\begin{equation}
\rho_n^2(e) = C \, q_n(e),
\label{eq:rho_factorized}
\end{equation}
where
\begin{equation}
q_n(e) \equiv \frac{g(n,e)}{n^2 S_n(n f_0)}
\label{eq:qn_def}
\end{equation}
encodes the relative harmonic power distribution, and $C$ is a normalization constant absorbing all eccentricity-independent factors.

Then we try to determine the constant $C$.
For each harmonic, according to Eq.~(\ref{eq:mean_2F}), the fully coherent $\mathcal{F}$-statistic satisfies
\begin{equation}
\langle 2\mathcal{F}_n \rangle = 2 + \rho_n^2 ,
\end{equation}
which motivates the definition
\begin{equation}
R_n \equiv 2\mathcal{F}_n - 2 ,
\end{equation}
with expectation value $\langle R_n \rangle = \rho_n^2$. 
For a given eccentricity $e$, the normalization factor $C$ can be estimated by matching the observed $\{R_n\}$ to the model prediction $\rho_n^2(e)$.
Here, for simplicity, we adopt an unweighted estimator for the normalization factor $C$, which is then profiled out in the subsequent weighted fit,
\begin{equation}
\hat C(e) =
\frac{1}{n_2-n_1+1}
\sum_{n=n_1}^{n_2}
\frac{R_n}{q_n(e)} .
\label{eq:N_estimator}
\end{equation}

Substituting $\hat C(e)$ back into Eq.~(\ref{eq:rho_factorized}), we construct a weighted least-squares statistic,
\begin{equation}
\chi^2(e)
=
\sum_{n=n_1}^{n_2}
\frac{\left[\hat C(e)\, q_n(e) - R_n\right]^2}
{4 + 4 R_n},
\label{eq:chi2_ecc}
\end{equation}
where the weights account for the variance of the non-central $\chi^2$ distribution of $2\mathcal{F}_n$, 
$\mathrm{Var}(2\mathcal{F}_n) \simeq 4 + 4R_n$.
The eccentricity estimate $\hat e$ is finally obtained by minimizing $\chi^2(e)$
\begin{equation}
\hat{e} = \underset{e}{\rm argmin}~\chi^2(e).
\end{equation}
The key advantage of this multi-harmonic eccentricity estimator is that it relies exclusively on the relative distribution of signal power among harmonics, rather than on the absolute signal amplitude.
As a result, the eccentricity is constrained primarily by the shape of $q_n(e)$ as a function of the harmonic index $n$, rendering the estimate largely insensitive to uncertainties in the intrinsic parameters.

Furthermore, this property significantly reduces the degeneracy between the eccentricity and other intrinsic parameters (such as $M$ and $s$) that commonly affect semi-coherent and fully-coherent searches.
By providing an accurate and robust pre-estimate of $e$, the method substantially reduces the effective parameter range explored in subsequent searches, thereby improving both the efficiency and robustness of the overall detection pipeline.

It is important to emphasize that although the current validation is performed using \ac{AK} waveforms, the underlying methodology is fundamentally waveform-independent. The multi-harmonic eccentricity estimation method proposed here is based on the relative power distribution among different harmonics. The core logic—exploiting the energy ratios across harmonics to constrain the orbital eccentricity—remains robust. This estimation framework is designed to be modular, ensuring that it can be readily adapted to more sophisticated relativistic waveforms as they become computationally viable.

\subsection{Particle swarm optimization}

The search for \ac{XMRI} signals is characterized by a high-dimensional parameter space ($D=10$) and a formidable computational burden. A traditional grid-based search is rendered prohibitive by the ``curse of dimensionality." To quantify this challenge, the total number of templates $N_T$ required to cover the parameter space $\Omega$ with a minimum \ac{FF} can be estimated using the metric geometry approach \cite{Cornish:2005hd,Gair:2004iv}:
\begin{equation}
N_T \approx \frac{1}{\left(2\sqrt{(1-FF)/D}\right)^D} \int_{\Omega} \sqrt{\det \Gamma_{\mu\nu}} d^D\lambda,
\label{eq:template_number}
\end{equation}
where $\det\Gamma_{\mu\nu}$ is the determinant of the metric tensor defined on the parameter space, and $D$ denotes the dimensionality. By evaluating the metric determinant at a representative point in the parameter space, we provide an order-of-magnitude estimate for the total number of templates. For a typical search targeting a maximum mismatch $ds^2 \le 0.03$ (equivalent to $FF \ge 0.97$) and a three-month signal, our preliminary estimation yields $N_T \sim \mathcal{O}(10^{42})$. This result is consistent with the order-of-magnitude estimates reported in Ref.~\cite{Moore:2019pke} for similar high-dimensional \ac{EMRI} searches.

Such an immense template bank makes brute-force grid searches computationally unfeasible. Moreover, the detection-statistic landscape for \ac{XMRI} is notably complex, featuring strong parameter correlations and a highly multimodal structure that further complicates the identification of the global optimum, as demonstrated in Appendix~\ref{app:f_degeneracy}. To efficiently navigate this landscape, we employ \ac{PSO} as the core global optimization engine in our search pipeline \cite{mohanty2018swarm,488968,EAbook}, allowing for an effective exploration of the parameter space without the prohibitive costs of a deterministic grid.

In the \ac{PSO} framework, a swarm of candidate solutions, referred to as particles, explores the parameter space. 
Each particle $i$ is characterized by a position $x_i$ and a velocity $v_i$, which evolve over successive iterations according to the following kinematic equations

\begin{subequations}
\label{eq:pso_update}
\begin{align}
\vv{v}_i^{k+1} &= w \vv{v}_i^k + c_1 r_1 (\vv{p}_{{\rm best},i} - \vv{x}_i^k)
\nonumber \\
&\quad+ c_2 r_2 (\vv{g}_{\rm best} - \vv{x}_i^k), \\
\vv{x}_i^{k+1} &= \vv{x}_i^k + \vv{v}_i^{k+1}.
\end{align}
\end{subequations}
where $\vec{p}_{{\rm best},i}$ denotes the best position historically attained by particle $i$, and $\vec{g}_{\rm best}$ represents the global best position discovered by the entire swarm. 
The coefficients $c_1$ and $c_2$ are the cognitive and social acceleration constants, respectively, while $w$ is the inertia weight designed to regulate the trade-off between global exploration and local refinement. 
The terms $r_1$ and $r_2$ are independent random variables sampled from a uniform distribution $U(0,1)$, introducing the stochasticity necessary to escape local sub-optima.

In our hierarchical search pipeline, \ac{PSO} is applied iteratively to maximize the semi-coherent or fully-coherent $\mathcal{F}$-statistic over the intrinsic source parameters at different stages of the analysis.
Each particle in the swarm represents a candidate point in the parameter space, and its fitness is evaluated by the corresponding value of the 
$\mathcal{F}$-statistic.
The optimization objective is therefore to locate the global maximum of  $2\mathcal{F}$, which corresponds to the maximum-likelihood estimate of the signal parameters after analytical maximization over the extrinsic parameters.

To handle the periodic nature of the angular parameters such as $\lambda_s,\beta_s,\iota,\alpha_0$ and ${\gamma}_0$, we implement periodic boundary conditions in the position update step \cite{mohanty2018swarm}. 
Furthermore, given the high dimensionality ($D=10$) and the multi-modal structure of the $2\mathcal{F}$ landscape, we employ a linear-decaying inertia weight strategy \cite{4223164,785511} to ensure that the swarm explores the entire parameter space extensively in the early stages and refines the search around the global maximum in the later stages.

\section{Implementation}\label{sec:implementation}

In this section, we describe the implementation of the proposed analysis framework.
We first generate and condition simulated TianQin TDI data, including a downsampling procedure designed to improve computational efficiency.
We then construct $\mathcal{F}$-statistic–based TDI templates using an efficient analytical approach, enabling high-throughput searches.
Finally, we introduce a hierarchical semi-coherent search pipeline, in which the coherent integration time is progressively increased to enhance the localization of the global maximum in parameter space.

\subsection{Data simulation and conditioning}

As the first step, we generate simulated TianQin data, which serve as the input for the subsequent search. 
The source parameters are specified as follows. 
For the central supermassive black hole, we adopt representative values of the mass $M$, luminosity distance $d_{\text{L}}$, and sky location $(\lambda_{\text{GC}}, \beta_{\text{GC}})$ consistent with existing observational constraints on Sgr~A* \cite{Ghez:2008ms,Gillessen:2008qv,EventHorizonTelescope:2022wkp}. 
The spin magnitude $s$ and its orientation $(\lambda_{s}, \beta_{s})$ are assigned according to the latest results from the Event Horizon Telescope \cite{EventHorizonTelescope:2022wkp,EventHorizonTelescope:2022wok}. 
The mass of the brown dwarf $m$ and the initial periastron distance $R_{\text{p}}$ are chosen to be representative of typical XMRI models discussed in the literature \cite{Amaro-Seoane:2019umn,Lin2022Identifying,Vazquez-Aceves:2022dgi,2025ApJ...987..208V}. 
The remaining angular parameters $\{\iota, \alpha_0, \gamma_0, \Phi_0\}$ are randomly drawn from their respective parameter ranges to ensure generality of the simulated signals. 

The total observing duration is set to 90 days.
For the source parameters considered in this work, this corresponds to a strong detection with a network signal-to-noise ratio of approximately $\ac{SNR}\simeq113$, which is already sufficient to enable precise constraints on the key source parameters.
Moreover, this duration naturally aligns with the ``3+3'' operating mode of TianQin, in which three months of continuous observation constitute a fundamental and self-contained analysis segment.
For consistency with this mission configuration and to focus on a single uninterrupted observing window, we therefore restrict our simulations and data analysis to the first three months of data in the ``3+3'' configuration.
Longer period of observation would naturally lead to better constraining ability.

Given these parameters, the simulated data are constructed through the following steps. 
First, the intrinsic \ac{GW} signal emitted by the XMRI system is computed, including its multi-harmonic structure and the slow secular evolution of the orbital parameters.
The signal is then projected onto the TianQin detector using the full time-dependent detector response, which accounts for the satellite orbital motion, antenna pattern, polarization response, and Doppler modulation.
Instrumental noise is added at the level of the single-link phase measurements, after which the time-delay interferometry observables are formed to obtain the $A$ and $E$ channels.
The resulting TDI time series, containing both instrumental noise and the injected \ac{XMRI} signal with parameters listed in Table~\ref{tab:source_parameters}, are initially recorded with a time resolution of $\Delta t = 0.5\,\mathrm{s}$ and are subsequently downsampled to a cadence of $100\,\mathrm{s}$ for both the fully coherent and semi-coherent analyses.
This corresponds to a Nyquist frequency of $0.005\,\mathrm{Hz}$, which captures the dominant harmonic content of typical \ac{XMRI} signals, with only extremely high-eccentricity sources contributing appreciable power near or above this frequency \cite{Amaro-Seoane:2019umn,2025ApJ...987..208V}.
The downsampled data, therefore, retain the relevant signal power while significantly reducing the computational cost of the search and validation analyses.

\begin{table}[htbp]
\renewcommand{\arraystretch}{1.3} 
\centering
\caption{Injected parameters of the simulated XMRI source and their physical meanings. The vector $\vec{S}$ denotes the directions of the spin of Sgr A*.}
\label{tab:source_parameters}
\begin{tabular}{l c l}
\hline
\hline
Symbol & Value & Description \\
\hline
$M \; (\mathrm{M_\odot})$               & $4 \times 10^6$   & Mass of Sgr A* \\
$m \; (\mathrm{M_\odot})$               & $0.05$            & Mass of the \ac{BD} \\
$s$                            & $0.9$             & Spin of Sgr A* \\
$e_0$                          & $0.7$             & Eccentricity at the initial time \\
$d_\text{L} \; (\text{kpc})$   & $8.0$               & Luminosity distance to Sgr A* \\
$R_\text{p} \; (R_\text{S})$   & $4.5$             & Initial periastron distance in\\
                               &                   & units of the Schwarzschild radius\\
                               &                   &($R_\text{S}=2GM/c^2$) \\
$\lambda_s \; (\text{rad})$    & $1.8$             & Ecliptic longitude of the spin vector $\vec{S}$ \\
$\beta_s \; (\text{rad})$      & $-0.3$            & Ecliptic latitude of the spin vector $\vec{S}$ \\
$\lambda_{\text{GC}} \; (\text{rad})$ & $4.6583$    & Ecliptic longitude of Sgr A* \\
$\beta_{\text{GC}} \; (\text{rad})$   & $-0.0977$   & Ecliptic latitude of Sgr A*\\
$\iota \; (\text{rad})$        & $0.2618$          & Inclination angle of the orbit \\
$\alpha_0 \; (\text{rad})$     & $1.0$             & Initial precession angle of \\
                               &                   &the orbital plane \\
$\gamma_0 \; (\text{rad})$     & $2.0$             & Initial argument of pericenter \\
$\Phi_0 \; (\text{rad})$       & $3.0$             & Initial orbital phase \\
\hline
\hline
\end{tabular}
\end{table}

\subsection{Efficient TDI Template Generation}\label{subsec:efficient_tdi}

The practical implementation of a large-scale search, such as the \ac{PSO} algorithm to be employed here, is fundamentally constrained by the computational cost of waveform generation. 
In our search configuration, each likelihood evaluation requires the generation of a three-month-long TDI template. 
To achieve a high-throughput search pipeline, we implement two key simplifications to accelerate the waveform calculation, focusing on the orbital dynamics and the detector response, respectively.

\subsubsection{Stationary eccentricity approximation}

For \acp{XMRI} observed over a relatively short duration ($T_c \leq 90\,\mathrm{days}$), the secular orbital evolution driven by radiation reaction is expected to be weak. 
In particular, the cumulative change in eccentricity over the observation time is well below the effective resolution of TianQin, which motivates treating the eccentricity as quasi-stationary, i.e.\ $\mathrm{d}e/\mathrm{d}t \simeq 0$. 
Adopting this approximation significantly simplifies the orbital dynamics and leads to a substantial reduction in waveform generation cost.

We have explicitly verified that this stationary-eccentricity approximation has a negligible impact on waveform fidelity within the scope of this work. 
For systems with moderate eccentricity, the mismatch between templates constructed with and without eccentricity evolution is $\lesssim 10^{-9}$, while even in the high-eccentricity regime (e.g.\ $e_0=0.93$), the fitting factor remains above $0.9921$. 
At the same time, the computational cost of template generation is reduced by a factor of $\sim 20$, making this approximation particularly advantageous for large-scale searches. 
We note, however, that waveform modeling for extremely high eccentricities remains intrinsically challenging, and the underlying waveform model itself becomes less accurate in this regime; this limitation is beyond the scope of this work and also does not qualitatively affect our conclusions.

\subsubsection{Analytic TDI response} \label{subsub_TDI}

The second computational bottleneck arises from the numerical evaluation of the \ac{TDI} response.  
Using \texttt{GWSpace} \cite{Li:2023szq}, which performs time-domain integration of the constellation motion and individual inter-satellite link responses, generating a three-month-long template for a single harmonic typically requires several seconds. 
For a representative system with initial eccentricity $e_0 = 0.7$, adequate signal fidelity requires $\sim 40$ harmonics. 
In a typical \ac{PSO}-based analysis, the detection statistic must be evaluated on the order of $\mathcal{O}(10^5)$ times to ensure reliable convergence \cite{Fu:2024cpu,mohanty2018swarm},  which would result in a total computational cost of order $\mathcal{O}(10^7)\,\mathrm{s}$ using the full numerical \ac{TDI} response, rendering direct evaluation computationally intensive.

To mitigate this challenge, we implement an analytic template construction based on a low-frequency approximation of the detector response. 
The main complication arises from relativistic orbital precession, which splits each harmonic $n f_0$ into multiple sidebands, making a full analytic treatment for each component complex. 
However, because the precession frequencies are typically small relative to the orbital frequency, these sidebands concentrate within a narrow frequency spread around each harmonic. Given this narrow bandwidth, the detector response remains nearly constant. We can therefore approximate the \ac{TDI} response $\mathcal{T}$ as constant across the narrow frequency width of each single harmonic cluster, allowing the detector response to be effectively decoupled from the precession-induced modulation. 
Mathematically, we decompose the $n$-th harmonic signal into a precession-modulated part $P(t)$, which is common to all harmonics and accounts for the evolution of angles $\alpha(t)$ and $\gamma(t)$, and a raw \ac{GW} phase component $H_n(t)$, such that
\begin{equation}
\mathcal{T}[P(t) H_n(t)] \approx P(t) \cdot \mathcal{T}[H_n(t)],
\end{equation}
where $\mathcal{T}[H_n(t)]$ is evaluated analytically at the central frequency of the $n$-th harmonic cluster. The complete mathematical formalism for integrating this analytic TDI response into the $\mathcal{F}$-statistic framework is detailed in Appendix~\ref{app:extrinsic}.

\begin{figure}[htbp]
    \centering
    \includegraphics[width=1.0\linewidth]{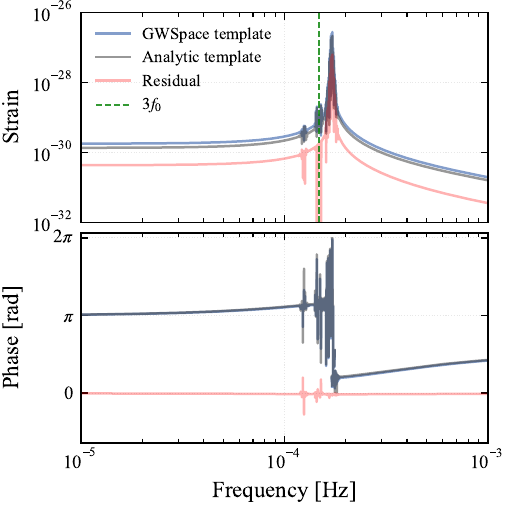}
    \caption{Comparison between the \texttt{GWSpace} template and the analytic template, obtained via direct Fourier transform of the time-domain signals. Upper panel: The strain of the $n=3$ harmonic for the \texttt{GWSpace} template (blue), the analytic template (grey), and their resulting strain residual (red). The vertical green dashed line marks the frequency $3f_0$, where $f_0$ denotes the orbital frequency of the \ac{XMRI} system, while the frequency sidebands surrounding the peak result from the combined effects of the \ac{XMRI} orbital precession and the periodic motion of TianQin. Lower panel: The phase evolution of the two templates and their corresponding phase residual (red). The close agreement between the templates, especially near the peak frequency, demonstrates the reliability of the analytic approximation used in our search pipeline.}
    \label{fig:approximate_waveform}
\end{figure}

This approximation substantially reduces the per-harmonic template generation time, from $3.08~\mathrm{s}$ for the full numerical approach to approximately $0.06~\mathrm{s}$ using the analytic method, as measured on a single core of Intel(R) Xeon(R) Gold 6330 CPUs (2.0 GHz). 
To verify that this efficiency does not compromise waveform accuracy, we compared the approximate templates against fully numerical \ac{TDI} waveforms across the target parameter space. 
As shown in Fig.~\ref{fig:approximate_waveform}, the agreement is high, with the fitting factor $\mathrm{FF} \simeq 0.998$ at a \ac{GW} frequency of $0.1~\mathrm{mHz}$, and $\mathrm{FF}$ further improves for higher-frequency harmonics, indicating the approximation is robust across the measurement band.

\subsection{Hierarchical semi-coherent search pipeline}
\label{subsec:pipeline}

\begin{figure*}[htbp]
    \centering
    \includegraphics[width=1.0\linewidth, trim={45pt 15pt 25pt 18pt}, clip]{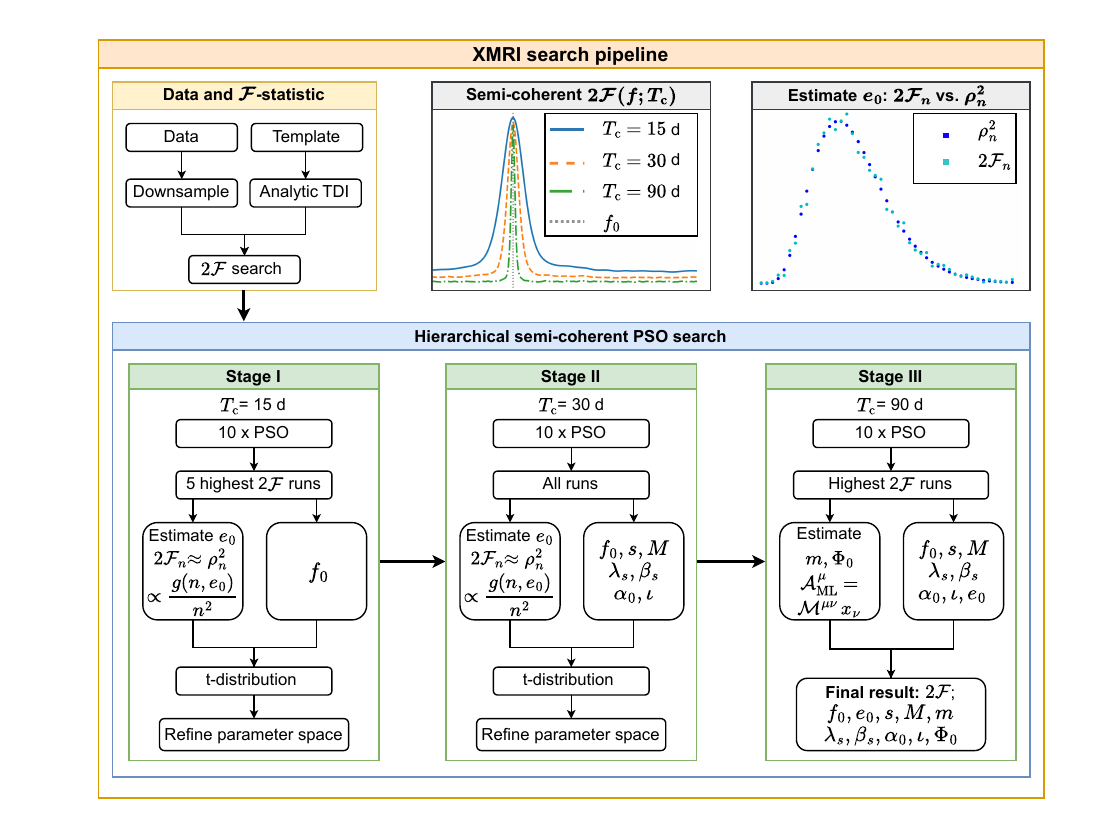}
    \caption{
    Schematic of the hierarchical semi-coherent search pipeline for \ac{XMRI} signals. 
    The top panels illustrate the foundational components: (left) the generation of the $\mathcal{F}$-statistic using downsampled data and analytic \ac{TDI} templates; (center) the evolution of the $2\mathcal{F}$ peak as the coherence time $T_c$ increases from 15 to 90 days. 
    The broader peaks at shorter $T_c$ provide a larger capture range that facilitates the initial global search, while the narrower peaks at longer $T_c$ yield significantly higher parameter precision, motivating our hierarchical semi-coherent approach; and (right) the use of the multi-harmonic power distribution $2\mathcal{F}_n$ to provide an initial probe of the initial eccentricity $e_0$. 
    The bottom panel details the three-stage hierarchical \ac{PSO} workflow. 
    Stage I ($T_c = 15$\,d) focuses on the initial localization of orbital frequency $f_0$ and initial eccentricity $e_0$. 
    Stage II ($T_c = 30$\,d) performs an intermediate refinement to break parameter degeneracies among the intrinsic parameters. 
    Stage III ($T_c = 90$\,d) executes the final fully coherent search to obtain high-precision estimates for the complete parameter set, including the analytically inferred extrinsic parameters. 
    At each transition, the parameter space is refined based on a $t$-distribution analysis of the best-performing \ac{PSO} runs.}
    \label{fig:pipeline}
\end{figure*}

Building on the simulated data and the efficient TDI template generation described above,
we now present the hierarchical semi-coherent search pipeline used to identify and characterize XMRI signals.
The pipeline is designed to balance detection sensitivity and computational efficiency,
and its overall structure is illustrated in Fig.~\ref{fig:pipeline}.
The search proceeds through three stages with progressively increasing coherence time,
allowing the parameter space to be iteratively refined.
We use $\Theta_i$ to denote the subset of parameters whose search ranges
are refined at Stage~$i$ within the hierarchical pipeline.

All stages of the search employ a \ac{PSO} algorithm with a fixed configuration.
Each PSO run uses a swarm of 100 particles and a maximum of 6000 iterations.
The cognitive and social acceleration coefficients are set to $c_1 = c_2 = 2$ following \citet{EAbook}. 
For the inertia weight $\omega$, while the literature suggests a linear decrease from 0.9 to 0.4, we have slightly tuned this range to 0.92--0.5 in our implementation to suit the XMRI search better.
To mitigate fluctuation raised by the stochastic evolution and to improve robustness, each semi-coherent stage consists of 10 independent \ac{PSO} runs executed in parallel.

We now turn to the computational aspects of evaluating the $\mathcal{F}$-statistic within each PSO iteration.
To reduce the computational cost while retaining the dominant signal power, the harmonic content included in the match filter is restricted adaptively based on the sampled orbital parameters and the sensitive band of the detector. 
For a given sampled orbital frequency $f_\mathrm{s}$, we first define the analysis harmonic range $[n_{f,\mathrm{min}}, n_{f,\mathrm{max}}]$, where $n_{f,\mathrm{min}} = \lceil 0.1\,\mathrm{mHz} / f_\mathrm{s} \rceil$ and $n_{f,\mathrm{max}} = \lfloor 5\,\mathrm{mHz} / f_\mathrm{s} \rfloor$, corresponding to the usable analysis band of the downsampled TianQin data, with the upper cutoff set by the Nyquist frequency. 
Simultaneously, for a sampled initial eccentricity $e_\mathrm{0,s}$, we utilize Eq.~(\ref{eq:snr_sum}) to identify the \ac{SNR}-dominant harmonic range $[n_{e,\mathrm{min}}, n_{e,\mathrm{max}}]$ that accounts for 99\% of the total squared \ac{SNR}. 
Consequently, for each \ac{PSO} particle, the effective range of harmonics evaluated in the likelihood is set as $[n_{\mathrm{min}}, n_{\mathrm{max}}]$, with $n_{\mathrm{min}} = \max(n_{f,\mathrm{min}}, n_{e,\mathrm{min}})$ and $n_{\mathrm{max}} = \min(n_{f,\mathrm{max}}, n_{e,\mathrm{max}})$. 
This dual-filtering approach ensures that only physically significant and detectable components are computed, thereby substantially reducing the computational cost of the likelihood evaluation.

\textbf{Stage I: 15-day semi-coherent search.} 
The first stage performs a semi-coherent search with a coherence time of $T_c = 15$ days. 
The \ac{PSO} explores the parameter set $\{f_0, e_0, s, M, m, \lambda_s, \beta_s, \iota, \alpha_0, \gamma_0\}$, with search range listed in Table~\ref{tab:pso_results}.
In this work, we restrict the search to systems with $e_0 \leq 0.9$, as higher eccentricities would produce hundreds of or even thousands of detectable harmonics, leading to a prohibitive computational cost.

Among the 10 \ac{PSO} runs, the five candidates with the highest values of the detection statistic $2\mathcal{F}$ are retained.
For each candidate, the corresponding orbital frequency $f_0$ is recorded, and an independent estimate of the initial eccentricity is obtained using the harmonic-based estimator.
Assuming approximate normality, we define the search range for the subsequent stage by constructing 99.7\% confidence intervals for $f_0$ and $e_0$ using a Student's $t$ distribution with $N_{\mathrm{s}}-1$ degrees of freedom, where $N_{\mathrm{s}}$ denotes the number of recovered parameter samples used in the estimation. 
This procedure accounts for the finite size of the available samples and provides a robust containment region for the refined search ranges over $\Theta_\mathrm{I} = \{ f_0, e_0 \}$.

\textbf{Stage II: 30-day semi-coherent search.} 
In the second stage, the coherence time is increased to $T_c = 30$ days.
Motivated by the weak sensitivity of the detection statistic to the \ac{BD} mass $m$ and the initial argument of pericenter $\gamma_0$ over the explored parameter space (Appendix~\ref{app:f_degeneracy}), these two parameters are fixed at the values that maximize $2\mathcal{F}$ in Stage~I.
The search ranges for the orbital frequency $f_0$ and the initial eccentricity $e_0$ are then restricted according to the statistical localization achieved in Stage~I.

The remaining parameters, excluding $m$ and $\gamma_0$, are refined using \ac{PSO} through 10 independent runs.
The resulting set of $f_0$ estimates, together with the corresponding eccentricity values $e_0$ inferred via the harmonic-based method, is treated as a statistical ensemble.
From this ensemble, $99.7\%$ confidence intervals are constructed using a Student's $t$ distribution.
These intervals define the localized search ranges for the subsequent stage over $\Theta_\mathrm{II} = \{ f_0, e_0, M, s, \lambda_s, \beta_s, \iota, \alpha_0 \}$.

\textbf{Stage III: 90-day coherent search.} 
The final stage of the pipeline consists of a fully coherent search with an integration time of $T_c = 90$ days. 
The search is carried out within the localized parameter ranges determined in Stage~II.
Within this framework, the candidate parameters that maximize the coherent $2\mathcal{F}$ statistic are identified as our best-fit estimates, representing the maximum-likelihood solution for the \ac{XMRI} signal. 
While the intrinsic parameters are directly optimized via \ac{PSO}, the remaining extrinsic parameters, namely the amplitude factor $h_0$ and the phase parameter, are analytically reconstructed from the maximum-likelihood estimators $\mathcal{A}_{\mathrm{ML},n}^{\mu}$ using the $\mathcal{F}$-statistic.
The physical parameters associated with these extrinsic quantities, in particular the \ac{BD} mass $m$, are then derived following the procedure outlined in Appendix~\ref{app:extrinsic}.
Given the demonstrated weak sensitivity of the detection statistic to the initial argument of pericenter $\gamma_0$ (Appendix~\ref{app:f_degeneracy}), and the fact that its effect is absorbed into the analytically resolved phase, we treat $\gamma_0$ as a nuisance parameter and do not further optimize it.
The true initial orbital phase $\Phi_0$ is recovered following the procedure described in Appendix~\ref{app:extrinsic}.

This hierarchical strategy enables efficient convergence of the search while maintaining high detection sensitivity and a manageable computational cost.
The computational cost of the search pipeline was assessed on a high-performance computing cluster using 50 cores of Intel(R) Xeon(R) Gold 6330 CPUs (2.0 GHz).
For a representative \ac{PSO} run with 100 particles and 6,000 iterations, corresponding to approximately $6\times10^{5}$ likelihood evaluations, the total wall-clock time was about 820 minutes ($\sim13.7$~h).

\section{Results}\label{sec:results}

In this section, we present the results obtained by applying the hierarchical semi-coherent search pipeline to simulated TianQin data.
\begin{table*}[t]
\caption{Summary of the five leading candidates from Stage I ($T_c = 15$~d), selected from 10 independent \ac{PSO} runs and ranked by the detection statistic $2\mathcal{F}$. The second and third columns list the injected parameters and the initial search ranges, respectively.}
\label{tab:pso_results}
\centering
\begin{tabular}{lcccccccccccc}
\hline\hline
Parameter & True Value & Search Range & PSO 1 (13913) & PSO 2 (13902) & PSO 3 (13701) & PSO 4 (13550) & PSO 5 (12883) \\
\hline
$f_0$ ($10^{-5}$~Hz) & 4.917112 & [1, 10]      & 4.91715 & 4.91717 & 4.91718 & 4.91692 & 4.91687 \\
$e_0$                & 0.7      & [0.01, 0.90] & 0.79652 & 0.72099 & 0.69891 & 0.71181 & 0.79627 \\
$s$                  & 0.9      & [0.01, 0.99] & 0.53476 & 0.82910 & 0.92055 & 0.83525 & 0.54612 \\
$M$ ($10^6 \mathrm{M_\odot}$) & 4        & [3.80, 4.20] & 4.08504 & 3.96717 & 3.92973 & 4.09793 & 4.02669 \\
$m$ ($\mathrm{M_\odot}$)      & 0.05     & [0.01, 0.09] & 0.02792 & 0.07387 & 0.01808 & 0.08223 & 0.06528 \\
$\lambda_s$ (rad)    & 1.8      & [0.01, 6.28] & 4.07126 & 1.60626 & 2.14005 & 1.54711 & 4.83483 \\
$\beta_s$ (rad)      & -0.3     & [-1.57,1.57] & 0.05889 & -0.04697 & -0.30035 & 0.55167 & 0.75279 \\
$\alpha_0$ (rad)     & 1        & [0.00, 6.28] & 4.14499 & 4.14171 & 0.98803 & 1.03185 & 3.91665 \\
$\gamma_0$ (rad)     & 2        & [0.00, 6.28] & 1.15388 & 3.17269 & 3.87818 & 2.62199 & 4.09951 \\
$\iota$ (rad)        & 0.2618   & [0.00, 3.14] & 0.33333 & 2.82616 & 0.40726 & 2.89655 & 0.49143 \\
\hline
\end{tabular}
\end{table*}

\subsection{Stage I} \label{sec:results_15day}

We first apply the hierarchical search pipeline to the simulated TianQin data using a coherent integration time of $T_c = 15$ days.
This initial stage aims to identify promising candidate regions in the multi-dimensional parameter space and to assess which source parameters can already be constrained at this coherence time.

The top five candidates, ranked by the semi-coherent $2\mathcal{F}$ statistic and obtained from ten independent \ac{PSO} runs, are summarized in Table~\ref{tab:pso_results}.
At this stage, the orbital frequency $f_0$ is consistently recovered with high precision, exhibiting a fractional error of order $\mathcal{O}(10^{-4})$ across all leading candidates.
In contrast, several other parameters—including the initial eccentricity $e_0$, Sgr~A* spin $s$, and mass $M$-remain weakly constrained.
This behavior is consistent with the strong parameter degeneracies inherent in the $\mathcal{F}$-statistic landscape, as discussed in Appendix~\ref{app:f_degeneracy}.

Despite these degeneracies, the recovered candidates exhibit a high degree of consistency in their inferred orbital initial eccentricity when analyzed using the multi-harmonic eccentricity estimator introduced in Sec.~\ref{sec:eccentricty_estimation}.
Applying this estimator to the five leading candidates yields the initial eccentricity estimates \[ e_{\mathrm{est}} = \{0.7003,\, 0.7007,\, 0.7015,\, 0.6983,\, 0.6872\}. \]
Based on a Student's $t$-distribution of these samples, we derive a refined search interval for the initial eccentricity at the $99.7\%$ confidence level \[ e_0 \in [0.6804,\, 0.7148], \] corresponding to a reduction of the effective eccentricity search range by nearly two orders of magnitude relative to the original one.
A similar refinement is obtained for the orbital frequency, with the updated interval \[ f_0 \in [4.9166,\, 4.9175] \times 10^{-5}\,\mathrm{Hz}. \]

These refined parameter ranges are adopted as the search ranges for the subsequent search stage with a longer coherent integration time.
By substantially restricting the explored parameter space, the initial 15-day semi-coherent search provides a well-localized starting point for the more computationally intensive refinement carried out in the subsequent stage of the pipeline.

\subsection{Stage II} \label{sec:results_30day}

In the second stage of the hierarchical pipeline, the coherent integration time is increased to $T_c = 30$ days, leading to improved parameter localization and a partial reduction of the dominant degeneracies identified in Stage I.

Table~\ref{tab:results_30d} summarizes the best-fit results obtained from Stage II. Compared to Stage I, the recovered values of $f_0$ and $e_0$ exhibit substantially reduced uncertainty, reflecting the combined effect of the longer coherent integration time and the refined eccentricity range. 
The intrinsic parameters associated with Sgr~A*, including the mass $M$ and spin $s$, are also more tightly constrained, although residual scatter remains.
Notably, while the search ranges for these parameters remained identical to those in Stage~I, their recovery precision is markedly improved. This is primarily attributed to the refined eccentricity range, which effectively breaks the parameter degeneracies between $e_0$ and the central black hole properties that previously hampered the search.

\begin{table}[htbp]
    \renewcommand{\arraystretch}{1.5}
    \centering
    \caption{Best-fit parameter recovery for Stage II ($T_c = 30$ d). The search ranges for the orbital frequency $f_0$ and initial eccentricity $e_0$ are refined based on Stage I, while the search ranges for all other parameters remain unchanged. The relative errors are calculated with respect to the injected parameters.}
    \label{tab:results_30d}
    \begin{tabular}{lccc}
        \hline\hline
        \makecell{Parameter\\~} & \makecell{Search \\Range} & \makecell{Best-fit\\Value} & \makecell{Relative\\Error} \\
        \hline
        $f_0 \; (10^{-5}\,\mathrm{Hz})$ 
            & [4.9166, 4.9175]   & 4.917120   & $1.6\times10^{-6}$ \\
        $e_0$                         
            & [0.6804, 0.7148]   & 0.7057     & $8.1\times10^{-3}$ \\
        $s$                         
            & [0.01, 0.99]   & 0.8987     & $1.4\times10^{-3}$ \\
        $M \; (10^{6} \mathrm{M_\odot})$      
            & [3.80, 4.20]       & 3.9110     & $2.2\times10^{-2}$ \\
        $\lambda_s\;(\mathrm{rad})$                 
            & [0.01, 6.28]       & 1.0525     & $4.2\times10^{-1}$ \\
        $\beta_s\;(\mathrm{rad})$                   
            & [$-1.57$, $1.57$]   & -0.0597    & $8.0\times10^{-1}$ \\
        $\alpha_0\;(\mathrm{rad})$                  
            & [0.00, 6.28]       & 1.0116     & $1.2\times10^{-2}$ \\
        $\iota\;(\mathrm{rad})$                     
            & [0.00, 3.14]       & 0.2660     & $1.6\times10^{-2}$ \\
        \hline\hline
    \end{tabular}
\end{table}

\begin{figure}[htbp]
    \centering
    \includegraphics[width=1.0\linewidth]{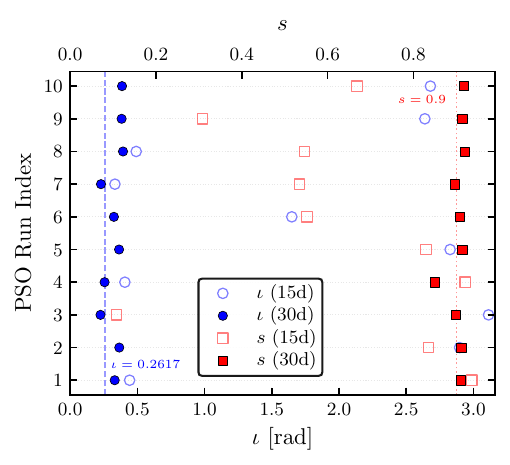}
    \caption{Recovered values of the orbital inclination angle $\iota$ (blue circles) and Sgr~A* spin $s$ (red squares) from ten independent PSO runs for Stage I (open) and Stage II (filled).
The results obtained in Stage II exhibit a substantially reduced scatter and cluster more closely around the injected parameters (vertical dashed lines), although a residual spread remains.}
    \label{fig:spin_iota_dual}
\end{figure}

The improvement in parameter recovery achieved in Stage II is further illustrated in Fig.~\ref{fig:spin_iota_dual}, which compares the distributions of the recovered orbital inclination angle $\iota$ and spin $s$ across independent searches for Stage I and Stage II.
For both parameters, Stage II results show a significantly reduced spread and a clear clustering toward the injected parameters, in contrast to the broader and more scattered distributions obtained at the shorter coherence time. This behavior reflects the combined impact of the refined eccentricity range, which mitigates the dominant parameter degeneracies, and the longer coherent integration time, which enhances the resolving power of the templates.

Despite this substantial improvement, residual scatter persists in the recovered values of $s$, $\iota$, and other correlated parameters, indicating that the search has not yet reached the precision expected from a fully coherent analysis of the full dataset.
Stage II, therefore, plays a crucial intermediate role in the hierarchical pipeline: it significantly refines the parameter space and stabilizes the recovery across independent searches, thereby providing a substantially narrower search range for the subsequent fully coherent stage to resolve the remaining degeneracies.

\subsection{Stage III}\label{subsec:stage_iii}

\begin{figure*}[htbp]
    \centering
    \includegraphics[width=1.0\linewidth]{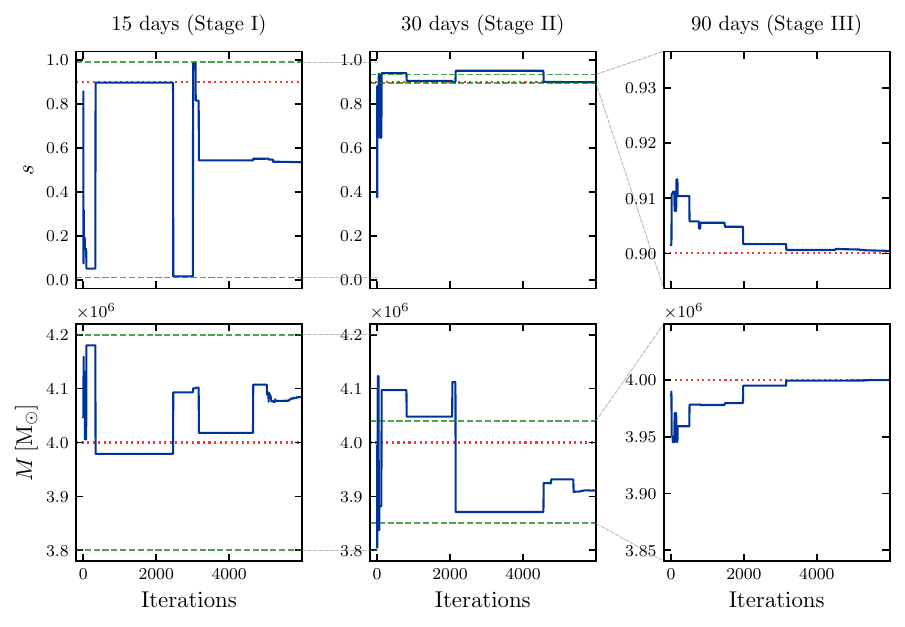}
    \caption{Multistage parameter localization for the Sgr A* spin $s$ (top) and mass $M$ (bottom) across the hierarchical pipeline. From left to right, the columns illustrate the progressive contraction of the search range and the corresponding enhancement in parameter precision as $T_c$ increases from 15 to 90 days. The blue solid curves show the PSO trajectories as a function of iteration number. The red dotted lines indicate the injected parameters, $s = 0.9$ and $M = 4\times10^{6}\,\mathrm{M_\odot}$. In Stages~I and~II, the green dashed lines represent the search ranges adopted for the subsequent stage: the bounds are inherited from the previous stage when no further restriction is applied, and are tightened once sufficient localization is achieved. Gray dashed connectors illustrate how these search ranges are propagated between successive stages. Gray dashed lines visualize the propagation of search ranges between successive stages. The convergence toward the injected parameters (red dotted lines) underscores the robustness of the strategy in mitigating parameter degeneracies through incremental coherence.}
    \label{fig:hierarchical_pso}
\end{figure*}

\label{sec:results_90day}

In the final stage of our hierarchical pipeline, the coherent integration time is extended to its maximum of $T_c = 90$ days.
The objective of this stage is twofold: to achieve the ultimate precision in the localization of intrinsic parameters within the previously constrained search range, and to analytically reconstruct the extrinsic parameters—including the \ac{BD} mass $m$ and the initial orbital phase $\Phi_0$—from the maximum-likelihood estimators.
This transition from a semi-coherent search to a fully coherent refinement allows us to break the remaining degeneracies, providing the final convergence required for high-precision parameter recovery.

\begin{table}[htbp]
    \renewcommand{\arraystretch}{1.5}
    \centering
    \caption{Best-fit parameter recovery for Stage III ($T_c = 90$~d). Intrinsic parameters are recovered via \ac{PSO} search, while $m$ and $\Phi_0$ are analytically reconstructed. The search ranges are refined from Stage II. The relative errors are calculated with respect to the injected parameters.}
    \label{tab:final_results}
    \begin{tabular}{lccc}
        \hline\hline
        \makecell{Parameter\\~} & \makecell{Search \\Range} & \makecell{Best-fit \\Value} & \makecell{Relative\\Error} \\
        \hline
        $f_0 \; (10^{-5}\,\mathrm{Hz})$ 
            & [4.91708, 4.91716] & 4.917102 & $2.0\times10^{-6}$ \\
        $e_0$                         
            & [0.6996, 0.6999]   & 0.6998   & $2.9\times10^{-4}$ \\
        $s$                         
            & [0.8984, 0.9194]   & 0.9005   & $5.6\times10^{-4}$ \\
        $M \; (10^{6} \mathrm{M_\odot})$      
            & [3.9014, 4.0789]   & 4.0001   & $2.5\times10^{-5}$ \\
        $\lambda_s\;(\mathrm{rad})$                 
            & [1.1054, 2.2332]   & 1.6584   & $7.9\times10^{-2}$ \\
        $\beta_s\;(\mathrm{rad})$                   
            & [$-0.6055, 0.3721$] & $-0.3751$ & $2.5\times10^{-1}$ \\
        $\alpha_0\;(\mathrm{rad})$                  
            & [0.9965, 1.0180]   & 1.0127   & $1.3\times10^{-2}$ \\
        $\iota\;(\mathrm{rad})$                     
            & [0.2541, 0.4061]   & 0.2679   & $2.3\times10^{-2}$ \\
        \hline
        $m \; (\mathrm{M_\odot})$ & $\cdots$ & 0.0532 & $6.4\times10^{-2}$ \\
        $\Phi_0 \; (\mathrm{rad})$ & $\cdots$ & 2.9892 & $3.6\times10^{-3}$ \\
        \hline\hline
    \end{tabular}
\end{table}

The recovered intrinsic parameters and analytically reconstructed extrinsic parameters from Stage III are summarized in Table~\ref{tab:final_results}.
Overall, the recovered values are consistent with the injected parameters within the estimated uncertainties.
For the intrinsic parameters, the primary quantities ($f_0, e_0, s, M$) are recovered with a precision of $\mathcal{O}(10^{-4})$ or better, with the orbital frequency reaching a relative error of $2.0\times10^{-6}$. The angular parameters ($\alpha_0, \iota$) are constrained to within a few percent. Notably, while the individual angular coordinates $\lambda_s$ and $\beta_s$ exhibit relatively high relative uncertainties—$7.9\%$ and $25\%$, respectively—the spin inclination angle (defined relative to the Galactic Anticenter) is constrained with a remarkably high absolute precision of $0.37^\circ$. This apparent discrepancy arises from the strong correlation and degeneracy between $\lambda_s$ and $\beta_s$ in the parameter space, a detailed visualization of this degeneracy is provided in Fig.~\ref{fig:2F_l_b} in Appendix~\ref{app:f_degeneracy}.
Regarding the extrinsic parameters, the analytical reconstruction yields $\hat{m} = 0.0532\,\mathrm{M_\odot}$ and $\hat{\Phi}_0 = 2.9892\,\mathrm{rad}$. The mass $m$ achieves a relative error of $6.4\times10^{-2}$, demonstrating that our weighted estimator effectively leverages information across multiple harmonics.
Compared to Stage I and Stage II, the dominant improvement achieved at this final stage is the substantial enhancement in parameter precision.

\begin{figure}[htbp]
    \centering
    \includegraphics[width=1.0\linewidth]{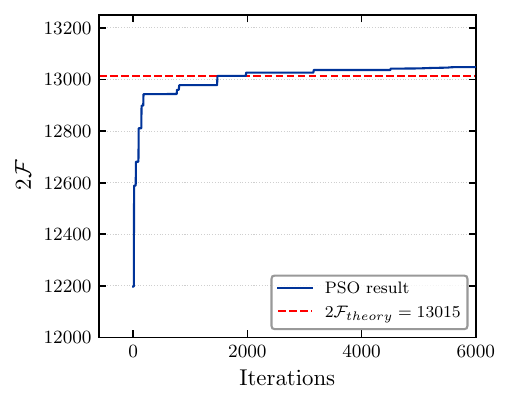}
    \caption{Evolution of the fully coherent $2\mathcal{F}$ statistic as a function of PSO iteration for the optimal chain in Stage III ($T_c = 90$~d). The solid blue curve represents the best-fit likelihood encountered by the swarm, while the red dashed line denotes the theoretical $2\mathcal{F}_{\mathrm{theory}}$ for the injected signal. The monotonic increase and subsequent saturation toward the expected value confirm that the PSO has successfully navigated the high-dimensional likelihood surface to the global maximum.}
    \label{fig:2F_iteration_90}
\end{figure}

The efficiency of the search in this high-precision regime is illustrated in Fig.~\ref{fig:2F_iteration_90}, which displays the evolution of the fully coherent $2\mathcal{F}$ statistic. The trajectory shows a rapid and stable climb toward the theoretical value, $2\mathcal{F}_{\mathrm{theory}} = 13015$, saturating at the global maximum within the first few thousand iterations. 
This unambiguous localization in Stage III indicates that the earlier-stage filtering has effectively eliminated secondary maxima that typically hinder fully coherent searches.

To further visualize the power of the hierarchical approach, Fig.~\ref{fig:hierarchical_pso} compares the parameter localization across all three stages. 
The progressive contraction of the search range for the spin $s$ and mass $M$ illustrates how incremental coherence and refined search range break the parameter degeneracies that are otherwise unresolved in Stage~I. 
While the initial stage provides a rough localization, the transition to Stage~III yields a dramatic sharpening of the parameter best fit values, localized precisely on the injected parameters (red dotted lines).

Stage III represents the final outcome of the hierarchical search.
It delivers parameter estimates with sufficient accuracy and provides a clear validation of the proposed semi-coherent strategy.
Moreover, the systematic improvement observed across increasing coherence times indicates that the method can be straightforwardly extended to even longer data sets, enabling high-precision inference for long-lived \ac{XMRI} signals observed by space-based gravitational-wave detectors.

\section{Discussion}\label{sec:discussion}

In this work, we have demonstrated that hierarchical semi-coherent searches provide a viable and robust framework for the detection of long-lived \ac{XMRI} signals. By progressively refining the search range, the method effectively overcomes the multimodal challenges of the likelihood surface, guiding the search toward the global maximum despite significant parameter degeneracies. 

The present analysis assumes relatively strong signals and isolated sources to focus on the intrinsic performance of the hierarchical strategy. In this regime, an accurate determination of the orbital eccentricity plays a crucial role, as it strongly constrains the harmonic structure and mitigates the dominant degeneracies that would otherwise compromise the fidelity of parameter recovery. Beyond the technical feasibility of detection, the high precision achieved in measuring the secondary mass $m$ --reaching a relative error of $\mathcal{O}(10^{-2})$ -- holds significant astrophysical implications. While the high precision reported here is obtained within the framework of \ac{AK} waveforms, these results primarily serve as a benchmark to validate the algorithmic efficiency of our hierarchical search strategy. Crucially, the proposed pipeline is intrinsically model-independent; as more faithful and computationally efficient relativistic waveforms—such as those derived from gravitational self-force theory—become available, they can be seamlessly integrated into this modular architecture.

Should a population of \ac{XMRI}s be detected in the future, such high-fidelity mass measurements would provide unprecedented constraints on the mass function of compact objects within the Galactic Center \cite{Amaro-Seoane:2019umn,Eckart_stellar}. This would enable us to refine dynamical evolution models of the nuclear star cluster \cite{2010RvMP...82.3121G} and probe the local stellar environment \cite{Genzel:2000mj}, including potential interactions with accretion disks or dark matter spikes surrounding Sgr~A* \cite{Gondolo:1999ef}. Moreover, the high-precision determination of the spin inclination angle to within an absolute error of $0.37^\circ$—offers a unique window into the angular momentum evolution of the Galactic Center, potentially revealing the alignment history between the supermassive black hole and its host galaxy \cite{Ciurlo:2025qns}.

The proposed framework is intrinsically scalable to longer observational baselines and can be straightforwardly extended to multi-year datasets without structural modifications.
The performance of our pipeline is primarily governed by the detectability of individual harmonics; specifically, the method remains viable as long as the $n$-th harmonic's frequency exceeds the $\sim 0.1$~mHz detection limit of space-based detectors and its corresponding \ac{SNR} surpasses the prescribed threshold of 7 \cite{LISA:2022yao}.
While this study focuses on moderately eccentric systems, the methodology is principally generalizable to higher eccentricities. The framework remains valid for systems with $e \lesssim 0.93$, a regime that ensures the stability of the stationary eccentricity approximation.
Despite these extreme cases, the underlying hierarchical search strategy remains a highly efficient framework for the high-precision characterization of the vast majority of expected, long-lived \ac{XMRI} signals within the Galactic Center. Finally, the high intrinsic \ac{SNR} of these sources suggests that the pipeline could be further extended to detect extragalactic \acp{XMRI} in nearby systems, such as the satellite galaxies of M82 at a distance of approximately 3.7~Mpc \cite{Amaro-Seoane:2020zbo}. For example, an \ac{XMRI} source that yields an \ac{SNR} $\sim 10^4$ at the Galactic Center (1,000 years before the plunge) would still maintain an \ac{SNR} $\sim 21$ at the distance of M82.

\section{Conclusion}\label{sec:conclusion}

In this work, we have developed a three-stage hierarchical semi-coherent pipeline for the search of long-lived \ac{XMRI} signals, specifically focusing on a brown dwarf orbiting Sgr~A*. Using simulated TianQin observations, we demonstrate that the pipeline robustly recovers the full set of parameters characterizing the system. Our results show that with a 90-day observation window, the intrinsic parameters ($f_0, e_0, s, M$) are determined with a precision of $\mathcal{O}(10^{-4})$ or better. Notably, the recovered Sgr~A* mass and spin magnitude achieve relative errors of $0.0025\%$ and $0.056\%$, respectively, while the spin inclination angle is constrained with an angular precision of $0.37^\circ$.

Furthermore, we show that the extrinsic parameters can be consistently reconstructed from the maximum-likelihood estimators, with the secondary mass $m$ achieving a relative error of $6.4\times10^{-2}$. This level of precision complements current electromagnetic constraints on the mass of Sgr~A* derived from stellar orbits~\cite{Ghez:2008ms,2019A&A...625L..10G} and significantly surpasses the resolution of spin estimates (both in magnitude and direction) obtained from horizon-scale imaging~\cite{EventHorizonTelescope:2022wkp}. Such gravitational-wave measurements provide a purely gravity-based probe of black hole properties, offering a unique and high-precision test of the Kerr metric in the strong-field regime without the systematic uncertainties inherent in astrophysical modeling of electromagnetic emissions.

In conclusion, hierarchical semi-coherent strategies offer a computationally efficient and reliable solution for analyzing long-duration signals in space-based gravitational-wave data. This approach is directly extensible to longer observing times and more complex systems, providing a foundation for high-precision galactic archaeology and fundamental physics tests in the era of space-borne detectors.

\begin{acknowledgments}
We thank Xian Chen, En-Kun Li, Jianwei Mei, Lorenzo Speri, Chang-Qing Ye and Jian-dong Zhang for useful discussions. We are also grateful to the anonymous referee for constructive comments and suggestions that improved the manuscript. This work has been supported by the National Key Research and Development Program of China (No. 2023YFC2206700), the science research grants from the China Manned Space Project (CMS-CSST-2025-A13), and the Fundamental Research Funds for the Central Universities, Sun Yat-sen University, Hebei Natural Science Foundation (No. A2023201041), Postdoctoral Fellowship
Program of CPSF (GZC20240366).

Generative AI was used to improve the clarity of language, and all outputs were reviewed and verified for accuracy by the authors.
\end{acknowledgments}

\section*{Data Availability}

The data that support the findings of this article are
openly available \cite{XMRIDataGitHub}.

\appendix

\section{\texorpdfstring{Mathematical Details of the $\mathcal{F}$-statistic Implementation}{Mathematical Details of the F-statistic Implementation}} \label{app:extrinsic}

In this appendix, we first present the explicit decomposition of the $\mathcal{F}$-statistic introduced in Eq.~\eqref{eq:factored_waveform} and subsequently describe the procedure for extracting the physical \ac{BD} mass $m$ and the initial orbital phase $\Phi_0$ from the multi-harmonic amplitude parameters recovered via the $\mathcal{F}$-statistic.

\subsection{\texorpdfstring{Analytical Formalism of the $\mathcal{F}$-statistic }{Analytical Formalism of the F-statistic}}

The data stream recorded by a space-based detector is the projection of the \ac{GW} signal onto the detector arms. Following Refs.~\cite{Cutler:1997ta, 2020PhRvD.102f3021H}, the dependence of the antenna-pattern functions $F_{+,\times}$ on the wave-frame orientation $\{\hat{n}, \psi\}$ can be factorized as
\begin{equation}
\begin{aligned}
F_+(t;\hat{n},\psi) &= a(t;\hat{n}) \cos 2\psi + b(t;\hat{n}) \sin 2\psi, \\
F_\times(t;\hat{n},\psi) &= b(t;\hat{n}) \cos 2\psi - a(t;\hat{n}) \sin 2\psi,
\end{aligned}
\label{antenna_pattern_function}
\end{equation}
where $\hat{n} = -\hat{k}$ is the unit vector pointing from the detector toward the source, and $\psi$ denotes the polarization angle.

For an eccentric \ac{XMRI} system, the detector response $h(t)$ is expressed as a superposition of multiple harmonics. Specifically, the $n$-th harmonic contribution $h_n(t)$ is obtained by projecting the waveform polarizations $h_{n,+,\times}$ [Eqs.~\eqref{xmri_sig}] onto the time-dependent antenna-pattern functions $F_{+,\times}(t)$:
\begin{equation}
h_n(t) = h_{n,+}(t;\theta,\phi) F_+(t;\hat{n},\psi) + h_{n,\times}(t;\theta,\phi) F_\times(t;\hat{n},\psi).
\label{eq:detector_response}
\end{equation}
For convenience, we define the following auxiliary coefficients:
\begin{equation}
\begin{aligned}
H_{n,0} &= \frac{\mu \omega_n^2}{d_{\mathrm{L}}} \frac{G}{c^4}, \\
A_{+,c} &= A_n(\sin^2 \phi - \cos^2 \phi \cos^2 \theta) \\
        &\quad + B_n(\cos^2 \phi - \sin^2 \phi \cos^2 \theta), \\
A_{+,s} &= -C_n \sin 2\phi (1 + \cos^2 \theta), \\
A_{\times,c} &= (A_n - B_n) \sin 2\phi \cos \theta, \\
A_{\times,s} &= -2 C_n \cos 2\phi \cos \theta.
\end{aligned}
\label{eq:f_stat_coefficient}
\end{equation}
Additionally, the total phase $\Phi_n(t)$ is separated into a time-evolving part $\Phi_{n,t}$ and a constant initial phase $\Phi_{n,0} = n\Phi_0$. By substituting this decomposition into Eq.~\eqref{eq:detector_response} and isolating the time-independent extrinsic parameters $\{H_{n,0}, \Phi_{n,0}\}$, we recast the detector response as a linear combination of the extrinsic amplitudes $\mathcal{A}_n^{\mu}$ and the intrinsic basis functions $h_{\mu,n}(t; \lambda)$. The extrinsic amplitudes are defined as
\begin{equation}
\begin{aligned}
    \mathcal{A}_n^1 &= H_{n,0} \cos \Phi_{n,0}, \\
    \mathcal{A}_n^2 &= H_{n,0} \sin \Phi_{n,0}, \label{eq:extrinsic_factor}
\end{aligned}
\end{equation}
while the explicit expressions for the harmonic components $h_{\mu,n}(t; \lambda)$ are given by
\begin{equation}
    \begin{aligned}
       h_{1,n}(t; \lambda)=&(A_{n,c}^{+}\cos2\psi-A_{n,c}^{\times}\sin2\psi)a(t;\hat{n})\cos\Phi_{n,t}\\
       +&(A_{n,c}^{+}\sin2\psi+A_{n,c}^{\times}\cos2\psi)b(t;\hat{n})\cos\Phi_{n,t}\\
       +&(A_{n,s}^{+}\cos2\psi-A_{n,s}^{\times}\sin2\psi)a(t;\hat{n})\sin\Phi_{n,t}\\
       +&(A_{n,s}^{+}\sin2\psi+A_{n,s}^{\times}\cos2\psi)b(t;\hat{n})\sin\Phi_{n,t},\\
       h_{2,n}(t; \lambda)=&(A_{n,s}^{+}\cos2\psi-A_{n,s}^{\times}\sin2\psi)a(t;\hat{n})\cos\Phi_{n,t}\\
       +&(A_{n,s}^{+}\sin2\psi+A_{n,s}^{\times}\cos2\psi)b(t;\hat{n})\cos\Phi_{n,t}\\
       -&(A_{n,c}^{+}\cos2\psi-A_{n,c}^{\times}\sin2\psi)a(t;\hat{n})\sin\Phi_{n,t}\\
       -&(A_{n,c}^{+}\sin2\psi+A_{n,c}^{\times}\cos2\psi)b(t;\hat{n})\sin\Phi_{n,t}.\\  
    \end{aligned}
\end{equation}

Following the approximation discussed in Sec.~\ref{subsub_TDI}, the \ac{TDI} response is treated as constant at the central frequency $f_n$ for the $n$-th harmonic. Within this framework, the first-generation Michelson $X$ combination is constructed by mapping the basis components $\{a(t)\cos\Phi_{n,t}, b(t)\cos\Phi_{n,t}, a(t)\sin\Phi_{n,t}, b(t)\sin\Phi_{n,t}\}$ to the \ac{TDI} variables $\{X^{(k)}\}_{k=1}^4$. Specifically, the first two components of the $X$ combination are given by
\begin{equation}
\begin{aligned}
\left[
\begin{array}{c}
X^{(1)} \\
X^{(2)}
\end{array}
\right] = & \left[
\begin{array}{c}
u_{2}(t) \\
v_{2}(t)
\end{array}
\right] \{ \text{sinc}[(1 + \hat{k}\hat{n}_{2})x/2] \cos[\Phi_{n,t} \\
& + (x/2)\hat{k}\vec{q}_{2} - 3x/2] + \text{sinc}[(1 - \hat{k}\hat{n}_{2})x/2] \\
& \times \cos[\Phi_{n,t} + (x/2)\hat{k}\vec{q}_{2} - 5x/2] \} \\
& - \left[
\begin{array}{c}
u_{3}(t) \\
v_{3}(t)
\end{array}
\right] \{ \text{sinc}[(1 + \hat{k}\hat{n}_{3})x/2] \cos[\Phi_{n,t} \\
& + (x/2)\hat{k}\vec{q}_{3} - 5x/2] + \text{sinc}[(1 - \hat{k}\hat{n}_{3})x/2] \\
& \times \cos[\Phi_{n,t} + (x/2)\hat{k}\vec{q}_{3} - 3x/2] \};
\end{aligned}
\label{eq:X12_definition}
\end{equation}
where the variables $X^{(3)}$ and $X^{(4)}$ are obtained by substituting the $\cos$ terms in Eq.~\eqref{eq:X12_definition} with $\sin$. The corresponding expressions for the $Y$ and $Z$ combinations are then derived through a cyclic permutation of the spacecraft indices following $1 \to 2 \to 3 \to 1$. 
The phase term $\Phi_{n,t}$ accounts for both the intrinsic frequency evolution of the $n$-th harmonic and the Doppler modulation induced by the detector's motion.
It is expressed as
\begin{equation}
\Phi_{n,t} = \omega_n t + \frac{1}{2}\dot{\omega}_n t^2 + (\omega_n + \dot{\omega}_n t) \frac{R}{c} \cos\beta \cos(\Omega t + \eta_o - \lambda).
\end{equation}
Here $\Omega = 2\pi/1~\text{yr}$, while $\eta_o$ denotes the initial ecliptic longitude at $t = 0$ of the detector, and \(R \equiv 1\,\mathrm{AU}\) represents the orbital radius around the Sun. The angular position of the source is specified by its ecliptic longitude $\lambda$ and latitude $\beta$. Detailed definitions of the remaining geometric factors and variables are provided in Ref.~\cite{Li:2023szq,Blaut:2009si}.

\subsection{Analytical recovery of extrinsic parameters}

Following the framework established in Sec.~\ref{subsec:F-statistic}, the log-likelihood $\ln\Lambda_n$ is maximized analytically with respect to the extrinsic amplitude parameters $\mathcal{A}_n^{\mu}$. Once the optimal intrinsic parameters are determined, the maximum-likelihood estimators for the extrinsic amplitudes are reconstructed via Eq.~\eqref{eq:ml_extrinsic_estimator}.

For the $n$-th harmonic, the two extrinsic amplitude parameters are obtained from Eq.~\eqref{eq:extrinsic_factor}. The overall amplitude factor $H_{n,0}$ can be recast by adopting the approximation $\mu \simeq m$ [Eq.~\eqref{reduced_mass}]:
\begin{equation}
H_{n,0}
= \frac{G\mu\, \omega_n^2}{d_{\mathrm{L}} c^4}
\simeq \frac{G\, m\, (2\pi n f_0)^2}{d_{\mathrm{L}} c^4}.
\end{equation}
From the two extrinsic amplitude parameters, we obtain
\begin{equation}
H_{n,0}^2 = (\mathcal{A}_{n}^{1})^2 + (\mathcal{A}_{n}^{2})^2 .
\end{equation}
This allows us to infer an estimate of the secondary mass from each harmonic,
\begin{equation}
m_n
= \sqrt{(\mathcal{A}_{n}^{1})^2 + (\mathcal{A}_{n}^{2})^2}\,  \frac{d_{\mathrm{L}}c^4}{G(2\pi n f_0)^2}.
\end{equation}
To optimally combine the information from multiple harmonics, we construct a
weighted average using the corresponding $\mathcal{F}$-statistic values,
\begin{equation}
\hat{m}
= \sum_n\frac{ 2\mathcal{F}_n}{\sum_n 2\mathcal{F}_n}\, m_n \,,
\end{equation}
which downweights harmonics with a lower signal-to-noise ratio.

The phase resolved directly from the extrinsic parameters represents a combined effect of the initial orbital phase $\Phi_{n,0}$ and an offset introduced by the argument of pericenter $\gamma_0$. This physical degeneracy is explored in depth in the subsequent Appendix~\ref{app:f_degeneracy}. In the $\mathcal{F}$-statistic framework, although the orbital phase $\Phi_{n,0}$ of each harmonic can be analytically recovered, it remains subject to a constant shift dependent on $\gamma_0$. To resolve this ambiguity, we utilize the phase difference between adjacent harmonics.

\begin{equation}
\Phi_{n+1,0} - \Phi_{n,0} = \Phi_0 .
\end{equation}
Using the recovered amplitude parameters, the phase difference for each adjacent pair is given by
\begin{equation}
(\Phi_0)_n
= \arctan\!\left[
\frac{{\mathcal{A}_{n}^{1}}{\mathcal{A}_{n+1}^{2}} -{\mathcal{A}_n^{2}}{\mathcal{A}_{n+1}^{1}}}
{{\mathcal{A}_n^{1}}{\mathcal{A}_{n+1}^{1}}+ {\mathcal{A}_{n}^{2}}{\mathcal{A}_{n+1}^{2}}}
\right].
\end{equation}
Finally, we combine the phase estimates from different harmonics using the same
$\mathcal{F}$-statistic weighting scheme,
\begin{equation}
\hat{\Phi}_0
= \sum_n\frac{2\mathcal{F}_n}{\sum_n 2\mathcal{F}_n}\, (\Phi_0)_n,
\end{equation}
yielding a robust estimator for the initial orbital phase.

\section{\texorpdfstring{Structure and degeneracies of the $\mathcal{F}$-statistic}{Structure and degeneracies of the F-statistic}}
 \label{app:f_degeneracy}

In this appendix, we examine the structure of the detection statistic $2\mathcal{F}$ in the parameter space of \ac{XMRI} to elucidate the challenges encountered during the stochastic search. The following numerical results provide the physical basis for the observed parameter insensitivity and degeneracies.

\subsection{Parameter insensitivity to $m$ and $\gamma_0$}

\begin{figure}[htbp]
    \centering
    \begin{minipage}{1.0\linewidth}
        \centering
        \includegraphics[width=0.85\linewidth]{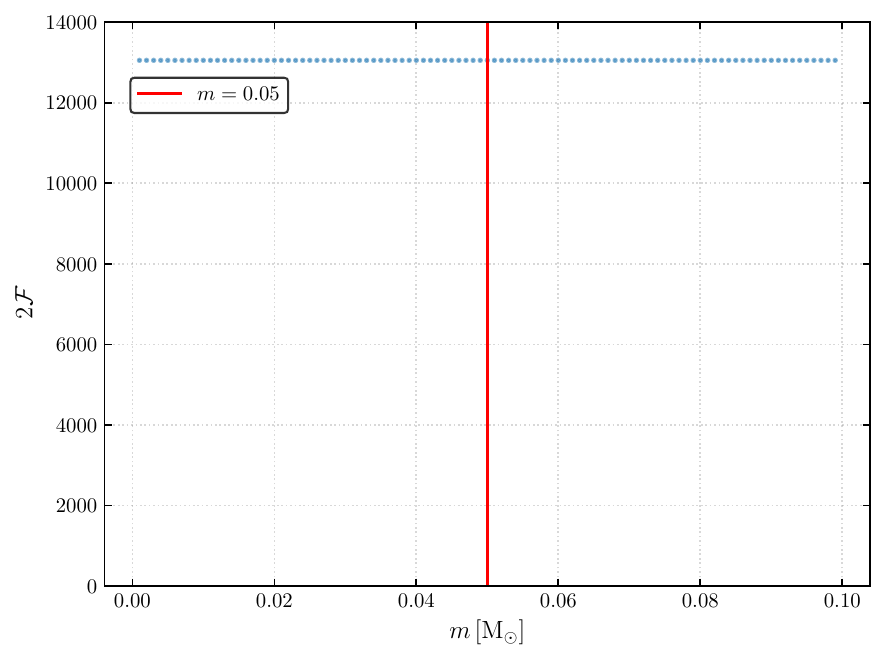}
        \vspace{-2mm} 
        \small{(a)}
    \end{minipage}
    
    \vspace{0.5cm} 
    
    \begin{minipage}{1.0\linewidth}
        \centering
        \includegraphics[width=0.85\linewidth]{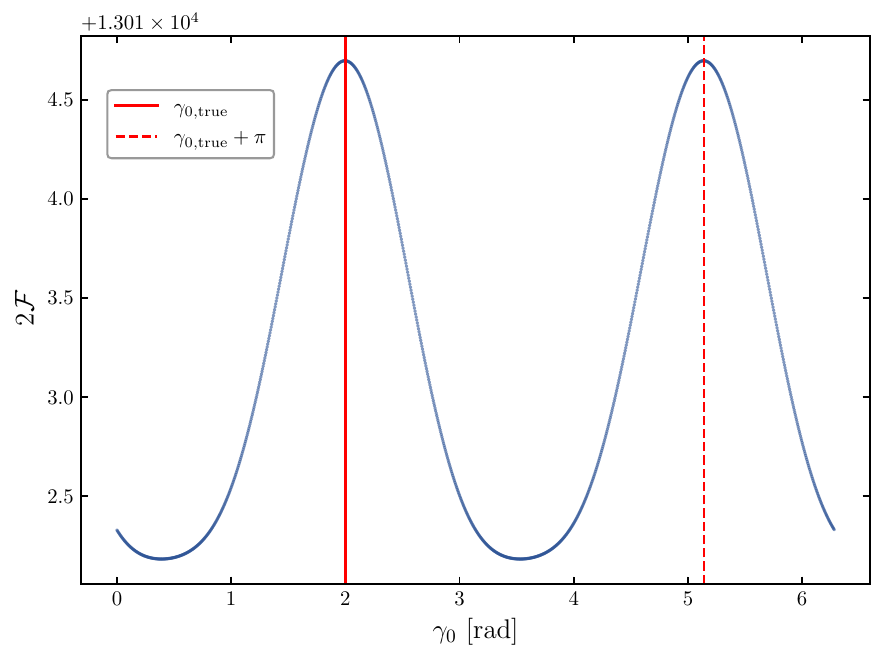}
        \vspace{-2mm}
        \small{(b)}
    \end{minipage}
    
    \caption{One-dimensional profiles of $2\mathcal{F}$ obtained by varying a single parameter while keeping all other parameters fixed at their injected parameters. The vertical red lines indicate the injected parameter values. (a) $2\mathcal{F}$ as a function of the \ac{BD} mass $m$, showing a flat distribution. (b) $2\mathcal{F}$ as a function of the initial argument of pericenter $\gamma_0$, exhibiting a secondary peak at $\gamma_{0,\mathrm{true}} + \pi$ resulting from the inherent phase-shift degeneracy.}
    \label{fig:2f_one_para}
\end{figure}

Fig.~\ref{fig:2f_one_para} presents the response of $2\mathcal{F}$ to variations in individual parameters while keeping other parameters fixed at their true values. 
As shown in Fig.~\ref{fig:2f_one_para} (a), the $2\mathcal{F}$ distribution is remarkably flat across the mass range. Physically, the \ac{BD} mass $m$ enters the signal evolution primarily through the frequency derivative $\dot{f}$.

Given our three-month observation window, the constraint on $\dot{f}$ remains weak, which directly leads to the observed insensitivity of the detection statistic to $m$.

In contrast, Fig.~\ref{fig:2f_one_para} (b) displays a primary peak at the true value and a prominent secondary peak at $\gamma_{0,\mathrm{true}} + \pi$, representing an inherent phase-shift degeneracy. Although peaks are present, the relative variation in $2\mathcal{F}$ is extremely small, typically on the order of $\mathcal{O}(1)$.
In a realistic data stream with Gaussian noise, such subtle variations can be easily obscured. Within the $\mathcal{F}$-statistic framework, variations in $\gamma_0$ are found to have only a marginal impact on the detection statistic, effectively rendering this parameter extrinsic-like.

Specifically, although $\gamma_0$ is not analytically partitioned into the extrinsic set in our formal derivation, the analytically resolved phase—obtained by maximizing the $\mathcal{F}$-statistic—effectively compensates for variations in $\gamma_0$, representing a combined contribution from $\Phi_{n,0}$ and $\gamma_0$.
This observation further motivates the tailored approach adopted in Appendix~\ref{app:extrinsic} for determining the true initial orbital phase $\Phi_0$ in the context of \ac{XMRI}.
Consequently, the impact of $\gamma_0$ is largely absorbed during the analytical maximization process, leading to the observed insensitivity of the $2\mathcal{F}$ surface to this parameter.

For other intrinsic parameters, the $2\mathcal{F}$ surface generally exhibits a complex multi-modal structure with numerous secondary peaks; however, these results are not presented here for the sake of brevity.

\subsection{Correlations and degeneracies of $e_0$ with $M$, $s$, and $\lambda_s$ with $\beta_s$}

\begin{figure}[htbp]
    \centering
    \begin{minipage}{1.0\linewidth}
        \centering
        \includegraphics[width=0.9\linewidth]{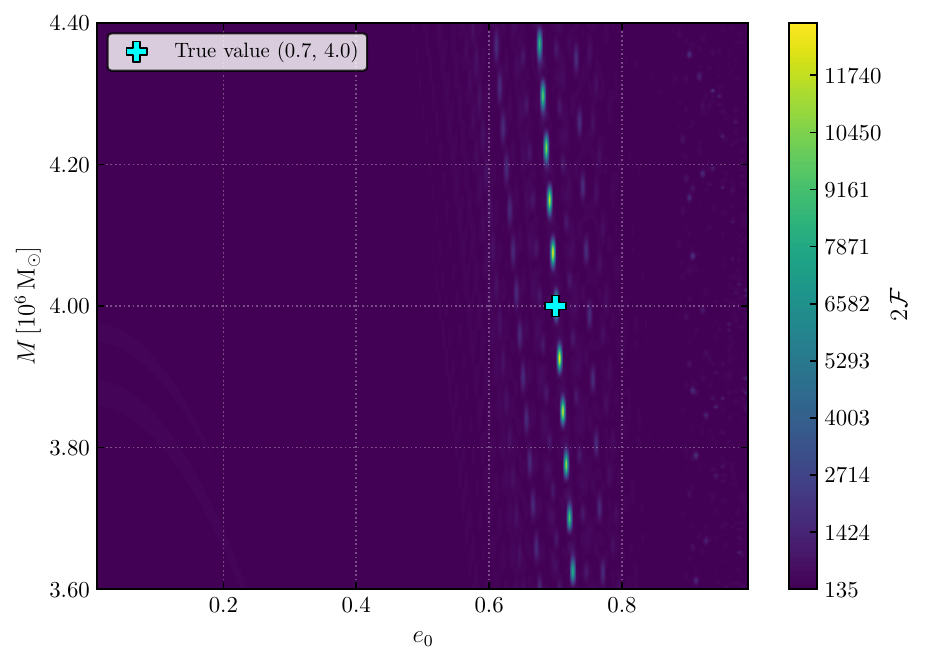}
        \vspace{-2mm}
        \small{(a)}
    \end{minipage}

    \vspace{0.6cm} 

    \begin{minipage}{1.0\linewidth}
        \centering
        \includegraphics[width=0.9\linewidth]{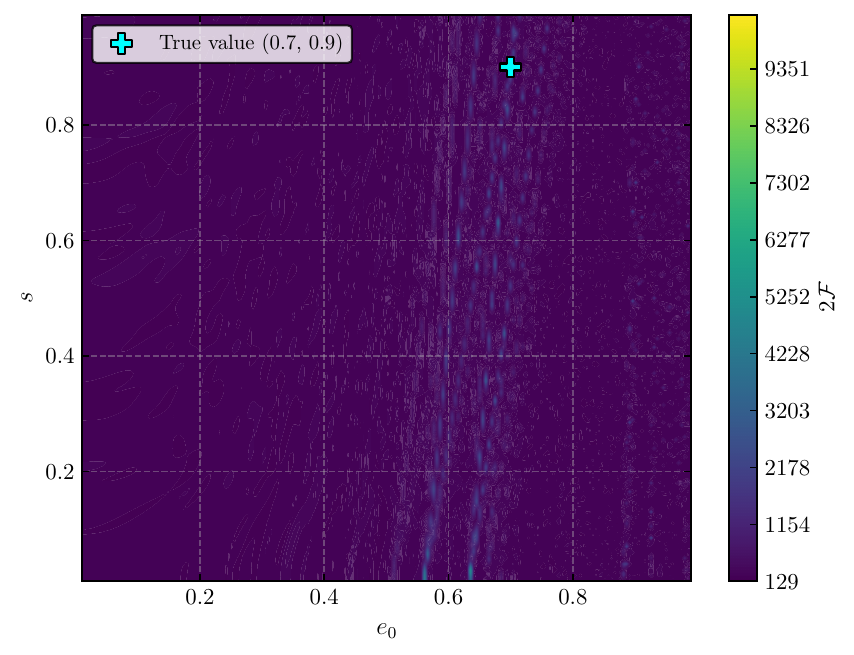}
        \vspace{-2mm}
        \small{(b)}
    \end{minipage}
    
    \caption{Two-dimensional $2\mathcal{F}$ surfaces evaluated over selected intrinsic-parameter subspaces. The surfaces are obtained by scanning the parameters shown on the axes, while keeping all remaining parameters fixed at their injected values. The cyan plus signs indicate the injected parameter locations. (a) $2\mathcal{F}$ surface in the $e_0$--$M$ plane. (b) $2\mathcal{F}$ surface in the $e_0$--$s$ plane.}
    \label{fig:2f_two_paras}
\end{figure}

Fig.~\ref{fig:2f_two_paras} illustrates the two-dimensional $2\mathcal{F}$ surfaces in the $e_0$-$M$ and $e_0$-$s$ planes, revealing strong correlations between the initial eccentricity and other intrinsic parameters.

As depicted in both panels of Fig.~\ref{fig:2f_two_paras}, the high-$2\mathcal{F}$ regions do not appear as isolated points but rather as elongated, tilted ridges. These ``degeneracy tracks'' imply that a deviation in $e_0$ can be largely compensated by corresponding shifts in Sgr~A* mass $M$ (see Fig.~\ref{fig:2f_two_paras} (a)) or spin $s$ (see Fig.~\ref{fig:2f_two_paras} (b)), maintaining a high statistic value.

We note that the maximum $2\mathcal{F}$ values in these surfaces are slightly lower than the theoretical expectation. This discrepancy arises from the finite resolution of the sampling grid, which may not perfectly align with the global maximum. Nevertheless, the surfaces in Fig.~\ref{fig:2f_two_paras} clearly demonstrate a complex multi-modal landscape with multiple secondary local maxima, necessitating robust global optimization strategies.

\begin{figure}[htbp]
    \centering
    \includegraphics[width=1.0\linewidth]{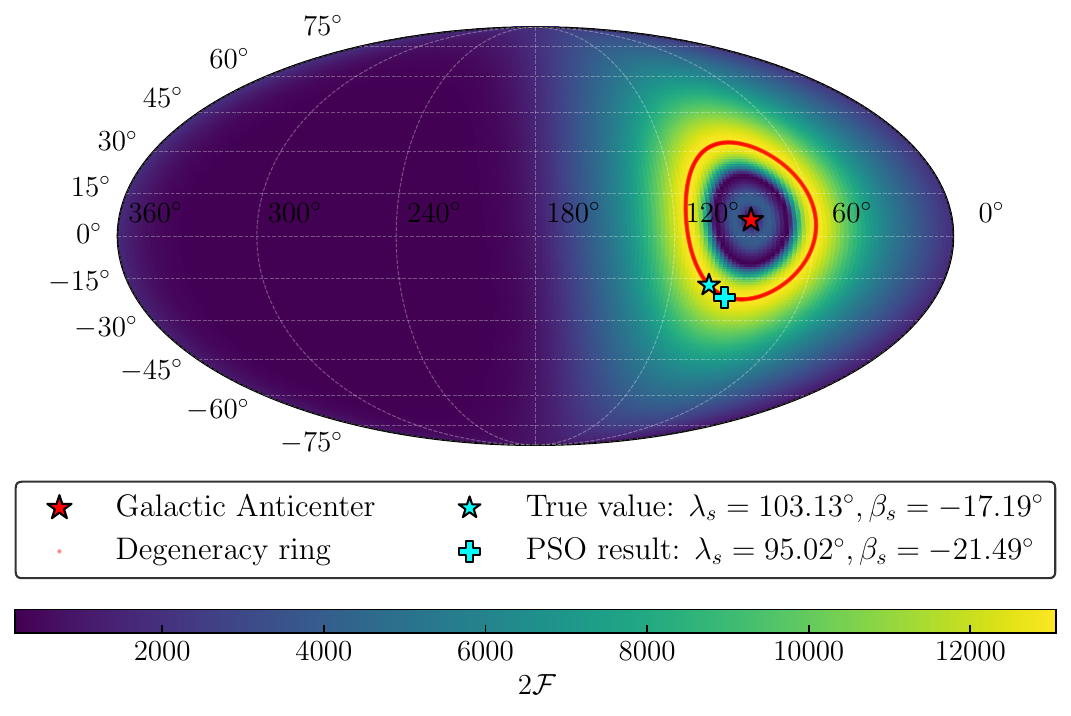}
    \caption{$2\mathcal{F}$ surface in the $\lambda_s$--$\beta_s$ plane, represented as a Mollweide projection in ecliptic coordinates. The surface is generated by varying the spin angular coordinates while keeping all other parameters fixed at their true values. The cyan star and the plus sign represent the injected parameters ($\lambda_{s}=103.13^{\circ}, \beta_{s}=-17.19^{\circ}$) and the best-fit \ac{PSO} candidate ($\lambda_{s}=95.02^{\circ}, \beta_{s}=-21.49^{\circ}$), respectively. The red star marks the location of the Galactic Anticenter. A prominent red solid line highlights the degeneracy ring, representing the locus of points that maintain a constant angular separation from the Galactic Anticenter, matching the opening angle of the true spin vector.
    }
    \label{fig:2F_l_b}
\end{figure}

Parallel to the correlations of $e_0$ with $M$ and $s$ discussed above, the spin orientation coordinates $\lambda_s$ and $\beta_s$ exhibit a strong mutual coupling.
As illustrated in Fig.~\ref{fig:2F_l_b}, the $2\mathcal{F}$ surface in the spin-angular subspace manifests as a prominent ring-like degeneracy centered on the Galactic Anticenter. 
Since the $2\mathcal{F}$ statistic remains consistently high along this trajectory, a stochastic search may recover various degenerate combinations of $\lambda_s$ and $\beta_s$ that reside on the ring. Consequently, while the individual coordinate uncertainties for $\lambda_s$ and $\beta_s$ appear relatively large, the spin inclination angle can be determined with high precision, as reported in Sec. \ref{subsec:stage_iii}.

\clearpage 
\bibliographystyle{apsrev4-2}
\bibliography{references_new}

\end{document}